\documentclass[aps,amsmath,superscriptaddress,amssymb,twocolumn,floatfix]{revtex4}     
\usepackage{bm}
\usepackage{graphicx}
\usepackage{amsmath}

\usepackage{color}

\newcommand{\hdad}{\mbox{HDA-d~}}

\newcommand{\comment}[1]{}

\begin{document}
\title{Liquid-liquid phase transition in simulations of ultrafast heating\\ and decompression of amorphous ice}

\author{Nicolas Giovambattista}
\email[Email: ]{ngiovambattista@brooklyn.cuny.edu}
\affiliation{Department of Physics, Brooklyn College of the City University of New York, Brooklyn, NY 11210, United States}
\affiliation{Ph.D. Programs in Chemistry and Physics, The Graduate Center of the City University of New York,\\ New York, NY 10016, United States}

\author{Peter H. Poole}
\email[Email: ]{ppoole@stfx.ca}
\affiliation{Department of Physics, St. Francis Xavier University, Antigonish, NS, B2G 2W5, Canada}
\date{\today}


\begin{abstract}
A recent experiment [K. H. Kim, {\it et al.}, Science {\bf 370}, 978 (2020)] 
showed that it may be possible to detect a liquid-liquid phase transition (LLPT)
 in supercooled water by subjecting high density amorphous ice (HDA) to ultrafast heating, 
after which the sample reportedly undergoes spontaneous decompression from a high density 
liquid (HDL) to a low density liquid (LDL) via a first-order phase transition. 
Here we conduct computer simulations of the ST2 water model, in which a LLPT is known 
to occur.  We subject various HDA samples of this model to a heating and decompression 
protocol that follows a thermodynamic pathway similar to that of the recent experiments.
Our results show that a signature of the underlying equilibrium LLPT can
 be observed in a strongly out-of-equilibrium process that follows this pathway 
 despite the very high heating and decompression rates employed here.
Our results are also consistent with the phase diagram of glassy ST2 water reported in
 previous studies.
\end{abstract}

\maketitle

\section{Introduction}

Starting in the 1970s, interest in the properties of water and amorphous ice expanded greatly in response to the seminal discoveries made by Austen Angell and coworkers on the anomalous properties of supercooled water~\cite{angell1,angell2}.  
Although the proposal of a liquid-liquid phase transition (LLPT) in supercooled water to explain these anomalies was 
first presented almost three decades ago~\cite{pooleNature}, 
experiments seeking to confirm and locate the LLPT remain 
challenging~\cite{nilssonNat2014,nilssonCpMax,kimmel2020,kimmel2021,coupin,katrin2011,suzukiMishimaGlycerol}.
The LLPT separates two distinct phases of liquid water, low-density liquid (LDL) and high-density liquid (HDL), 
which are proposed to be the liquid-state manifestations of the two experimentally known forms of amorphous ice,
 low-density amorphous (LDA) and high-density amorphous (HDA)
 ice (see, e.g., 
Refs.~\cite{mihsimaStanley1998,stanleyDebenedettiPhysToday,thomasReview,angellAmorphIce,nilssonReview,galloReview,pablo2003,handleFrancescoRev}).  
The estimated location of the LLPT in the phase diagram of water places it at deeply supercooled 
temperatures for the liquid phase and at elevated pressures in the kbar 
range~\cite{mishima1,mihsimaStanley1998,mishima2010,nillsonShift,skinnerRev}. 
 Under these conditions, the supercooled liquid rapidly transforms to the stable ice phase, 
and so an experiment to detect the LLPT must measure properties of the metastable liquid 
state on a time scale long enough to capture the behavior of the liquid in equilibrium, 
but do so fast enough to complete the measurement before ice nucleation eliminates 
the liquid phase~\cite{pabloFrancescoReview}.

A recent experimental study~\cite{katrinNilsson} addressed these challenges using a novel method to prepare HDL under pressure via ultrafast ($\sim100$~fs) isochoric heating of HDA using an IR laser pulse.  The HDL sample so created then relaxed to ambient pressure on a \mbox{sub-$\mu$s} time scale, triggering the phase transition to LDL, followed by crystallization on a longer time scale of approximately $10~\mu$s.  Short ($<50$~fs) X-ray laser pulses were used to measure the time evolution of the structure factor during the relaxation of the HDL sample to ambient pressure, and revealed the conversion of the sample to LDL prior to crystallization.

In the present work, we use computer simulations to study the thermodynamic pathway explored in Ref.~\cite{katrinNilsson} 
in order to clarify the response of the system when subjected to such rapid changes of state.  For our simulations we use the ST2 model of water, in which the occurrence of a LLPT is well 
documented~\cite{pooleNature,pabloNature,smallenburgST2,liu}.  
As a consequence, we study the case where we know that the equilibrium liquid system exhibits a LLPT, and so we can compare the thermodynamic pathway followed in the experiment to the known location of the LLPT in ST2 water, and we can test how the LLPT manifests itself along this pathway.

Our simulation results verify that rapid isochoric heating of ambient-pressure HDA drives the sample into the HDL region of the phase diagram at pressures above the coexistence line of the LLPT.  
Our simulations also reproduce the signature of the LLPT during the relaxation of the sample back to ambient pressure.
To provide broader context, we explore a wide range of paths similar to that used in Ref.~\cite{katrinNilsson} by varying both the density of the initial HDA sample, and the temperature to which it is heated before it is relaxed back to ambient pressure.
We find that even when equilibrium liquid states of HDL and LDL do not have time to be established during this process, the signature of the underlying HDL-LDL transition can be observed in the response of the system.
Overall, our simulations confirm the interpretation of the thermodynamic pathway explored in Ref.~\cite{katrinNilsson} and demonstrate that ultrafast manipulation of samples of amorphous ice and supercooled water can be used to confirm the existence, and also find the location, of the LLPT in the phase diagram of water.

\begin{figure}      
\centerline{
\includegraphics[width=7cm]{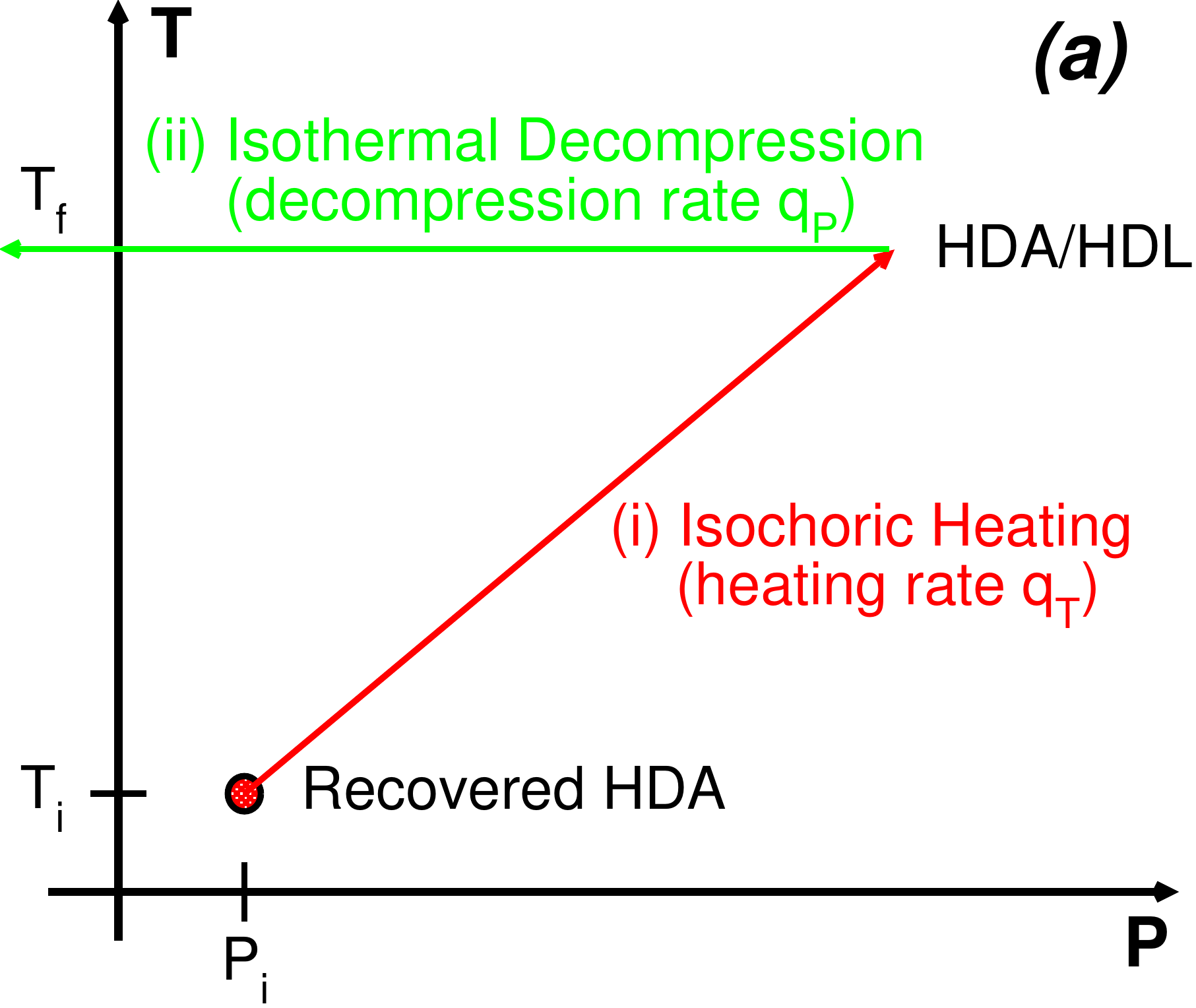}
}
\bigskip
\centerline{
\includegraphics[width=7cm]{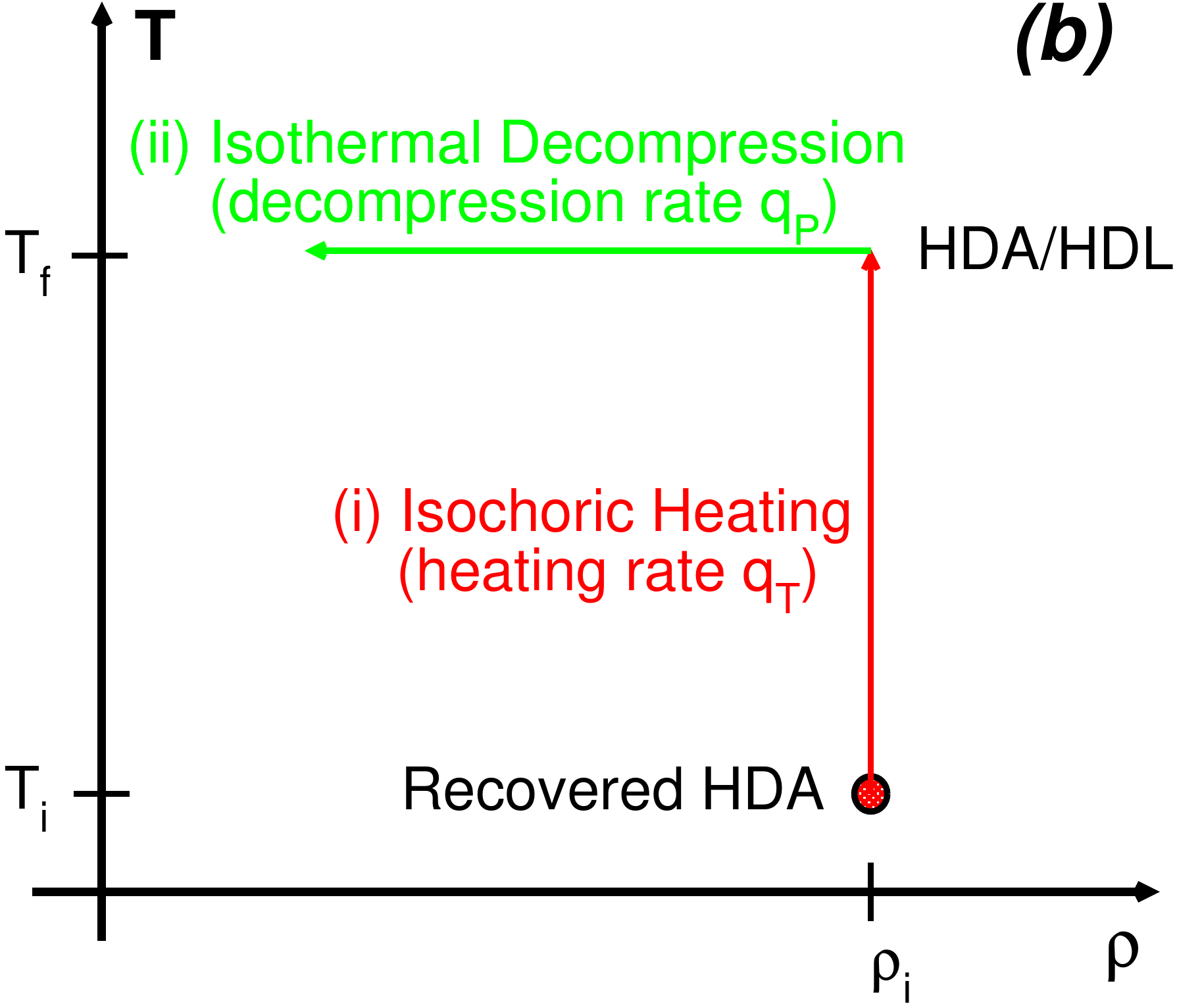}
}
\caption{Schematic diagrams 
showing the paths in the (a) $T$-$P$ and (b) $T$-$\rho$ planes followed during our computer
 simulations, which mimic the experiments of Ref.~\cite{katrinNilsson}. 
An HDA sample is recovered at pressure $P_i$ and temperature $T_i$ with density $\rho_i$. 
 The recovered HDA sample is then heated at constant density 
at rate $q_T$ up to a temperature $T_f$. Depending on the value of $T_f$, the sample may be found 
 in the HDA or HDL state. The HDL/HDA so produced is then
 decompressed  at constant temperature $T_f$.  In our computer simulations,
$T_i=80$~K and various values of $T_f$ and $P_i$ are explored ($T_f=100$ to $300$~K, $P_i=-100$ to $1150$~MPa). In the experiments of  Ref.~\cite{katrinNilsson}, 
$T_i=115$~K, $T_f \approx 205$~K, and $P_i \approx 0$~MPa.
}
\label{schemeProcess}
\end{figure}

\section{Experimental pathway}
\label{introSec}

In the experiments of Ref.~\cite{katrinNilsson}, 
HDA samples are produced under pressure and then recovered under vacuum (pressure $P \approx 0$) at temperatures $T$ in the range $T=78$ to $115$~K.
These HDA samples are then subjected to a process of rapid heating followed by density relaxation along the thermodynamic path depicted in Fig.~\ref{schemeProcess}.  

The details of this procedure are as follows.  Prior to the heating process, the HDA samples are at an initial pressure of $P_i\approx 0$ and initial temperature $T_i=115$~K.  Under these conditions, Ref.~\cite{katrinNilsson} estimates that the initial density $\rho_i$ of the HDA sample is in the range $1.13$ to $1.16$~ g/cm$^3$.
The HDA sample is then rapidly heated from $T_i$
to a final temperature $T_f=205 \pm10$~K using an infrared (IR) laser pulse that lasts for approximately
 $100$~fs. The corresponding heating rate was estimated to be $q_T \simeq 4500$~K/ns~\cite{katrinNilsson}.
As explained in Ref.~\cite{katrinNilsson}, 
this heating process occurs much faster than the rate, governed by the speed of sound in the sample, required for the sample density to respond to the change in $T$.   As a result, the volume of the sample remains approximately constant and so
the heating process is isochoric.

Since the heating process is isochoric, immediately after the IR pulse is applied, the sample density is still $\rho_i$ but the sample temperature is now $T_f \approx 205$~K.  Based on experimental data for supercooled liquid water, Ref.~\cite{katrinNilsson} shows that a liquid sample at this density and temperature would be found at $P=300\pm 50$~MPa and have a liquid-state relaxation time of less than 10~ns.  Therefore, within a few ns of the IR heating process, the sample is in the equilibrium HDL state at $T \approx 205$~K and $P\approx 300$~MPa.  However, the surface of the sample is exposed to ambient pressure conditions ($P \approx 0$), and so the sample density immediately begins to decrease as the system relaxes toward the lower equilibrium density expected at $P\approx 0$.  In Ref.~\cite{katrinNilsson} this density
 relaxation is described as a ``decompression" process because the pressure inside the sample is decreasing from a high value, transiently imposed by the IR pulse, to $P \approx 0$.  
The sample temperature 
remains constant during the density relaxation for approximately $100~\mu$s, until heat conduction
 from the surroundings begins to cool the sample.
That is, during the first $100~\mu$s, the density relaxation process is analogous to an isothermal
 decompression. 

During this isothermal decompression, the structure of the sample is studied 
using short ($<50$~fs) X-ray pulses over a time window of 
$8.4$~ns to $1$~ms after the IR pulse is applied.
From the evolution of the structure factor with time, Ref.~\cite{katrinNilsson} shows that the sample converts from HDL to LDL during the isothermal decompression, before crystallization intervenes on a time scale of approximately $10~\mu$s.
  Since the pressure inside the sample is not measured in these experiments, it is not clear what 
the decompression rate $q_P$ is.  However, it is estimated that the 
HDL-to-LDL transformation occurs within $10-100$~ns
after the IR pulse~\cite{katrinNilsson}. 
In the Supplementary Material of Ref.~\cite{katrinNilsson} it is 
estimated that the decompression process lasts for $70$~ns, 
giving an experimental decompression rate of approximately 
$q_P = 4.3$~MPa/ns. 

\begin{figure}      
\centerline{
\includegraphics[width=7cm]{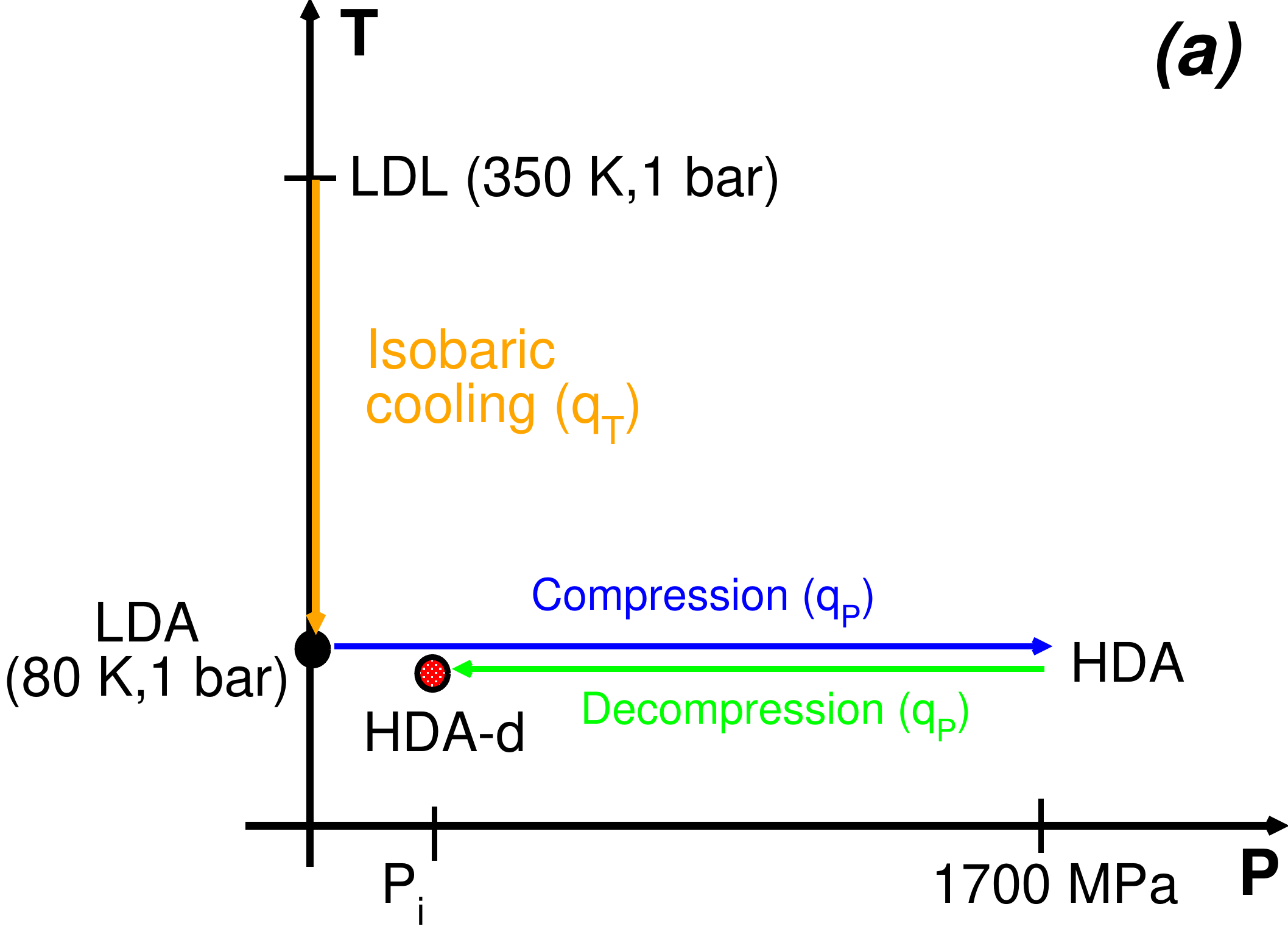}
}
\bigskip
\centerline{
\includegraphics[width=7cm]{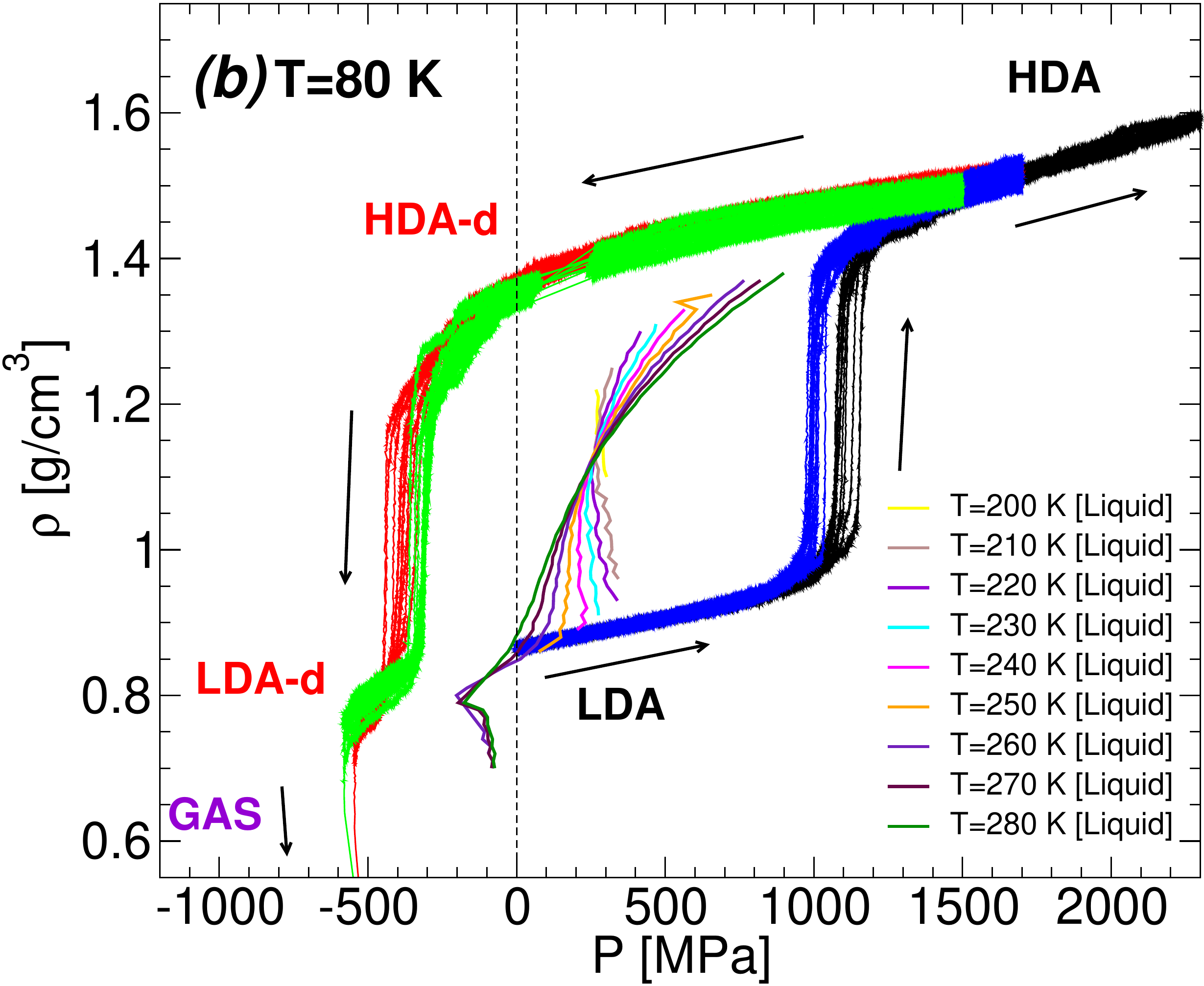}
}
\caption{(a) Schematic diagram showing the process followed in our simulations
to prepare HDA-d. 
(b) Density as function of pressure during the pressure-induced 
LDA-to-HDA transformation and subsequent HDA-to-LDA transformation at $T=80$~K
[right-arrows and left-arrows in (a)]. 
{\color{black}Black and red lines are 
obtained with a compression/decompression rate $q_P=300$~MPa/ns}; blue and
green lines correspond to $q_P=30$~K/ns. All $10$ independent runs are included.
The sharp density changes at $P\approx 1000$ and $1100$~MPa indicate the compression-induced LDA-to-HDA transformation.  
The sudden changes in density at $P\approx -350$ and $-400$~MPa signal the decompression-induced HDA-to-LDA transformations.  
Recovered LDA samples fracture at $P \approx -550$~MPa.
For comparison, we also include the isotherms of the
 equilibrium liquid at temperatures below and above the LLCP 
temperature $T_c\approx 245$~K ($\rho_c \approx 0.955$~g/cm$^3$) from Ref.~\cite{poole2005}.
Reducing the compression/decompression rate $q_P$ 
reduces the hysteresis during the 
LDA-HDA transformation, bringing the LDA-to-HDA transformation paths 
(red and green lines) as well as the HDA-to-LDA transformation paths 
(black and blue lines) closer to the low-temperature liquid isotherms. 
}
\label{HDALDALIQ-Rates}
\end{figure}

\section{Simulation Methods}
\label{simuls}

In this work, we conduct out-of-equilibrium molecular dynamics simulations 
of a system composed of $N=1728$ water molecules in a cubic box with periodic 
boundary conditions. We employ the ST2 water model~\cite{ST2model}, 
which exhibits a LLPT 
and associated liquid-liquid critical point (LLCP) in the supercooled 
liquid domain.  The critical values of the temperature, pressure and density for the LLCP in ST2 are respectively
$T_c= 247\pm3$~K, $P_c= 185\pm15$~MPa, 
and $\rho_c=0.955\pm0.01$~g/cm$^3$~\cite{pooleNature,francisPooleST2dynam,liu,cuthbertson,pabloFrancescoReview,pabloNature}. 
The ST2 water model has been used extensively 
to study the behavior of water in the 
liquid~\cite{pooleNature,pabloFrancescoReview,smallenburgST2,liu} and 
glassy states~\cite{poole1993,MySciRep,chiu1,chiu2,ourST2PEL-1,ourST2PEL-2,ourST2PEL-3}.
For consistency with previous studies, we treat the long-range electrostatic interactions
 using the reaction field technique~\cite{allen}.  Refs.~\cite{chiu1,chiu2,poole2005} provide reference data sets for the ST2 
system employed here.  In all simulations, $T$ and $P$ are
controlled using a Berendsen thermostat and barostat, respectively.

To carry out cooling or heating runs, we change the thermostat temperature
linearly with time at a rate $q_T$.  To conduct compression or decompression
runs, the barostat pressure changes linearly at a rate $q_P$; see Ref.~\cite{chiu1} for details.
For each cooling, heating, compression, or decompression study, we generate $10$
independent starting configurations using the same protocol for sample
preparation.  These $10$ configurations are then subjected to the
chosen process, allowing us to assess the
magnitude of the sample-to-sample variation due to differences in the starting configurations.

Our starting samples of HDA are created via the procedure described in detail 
in Refs.~\cite{chiu1,chiu2,ourST2PEL-1,ourST2PEL-2,ourST2PEL-3} and illustrated schematically in Fig.~\ref{HDALDALIQ-Rates}(a).
We first produce LDA by cooling the equilibrium liquid from $T=350$~K
 to $T=80$~K at constant $P=0.1$~MPa using a cooling rate $q_T=30$~K/ns.
This LDA sample is then compressed isothermally at $T=80$~K from $P=0.1$~MPa to 
$P=1700$~MPa using a compression rate 
$q_P=300$~MPa/ns.
Fig.~\ref{HDALDALIQ-Rates}b shows the density of $10$ independent simulations 
during the compression of LDA.
The sharp change in density at 
$P \approx 1100$~MPa indicates 
the transformation of LDA to HDA.
The HDA samples produced by this transformation are then decompressed at the 
same $T=80$~K and rate $q_P$ used during the compression of LDA.  
The density of HDA during the decompression process is shown in
 Fig.~\ref{HDALDALIQ-Rates}b.
The sharp density change at 
$P \approx -400$~MPa 
indicates the conversion of HDA back to LDA~\cite{chiu1,ourST2PEL-1}. 
The sudden decrease in 
density at $P \approx -550$~MPa corresponds to 
the limit of stability of LDA at which the sample fractures.
In this work, HDA samples produced by decompressing HDA to pressures below $P=1100$~MPa are denoted as HDA-d.

For the present simulations, we seek to study heating-decompression processes similar to that realized in the experiments of 
Ref.~\cite{katrinNilsson} and described in Sec.~\ref{introSec}.  This pathway starts with a sample of HDA at an initial temperature $T_i$ and initial pressure $P_i$, corresponding to an initial density $\rho_i$, as depicted in 
Fig.~\ref{schemeProcess}.  This sample is heated isochorically to a final temperature $T_f$, followed by an isothermal decompression carried out at $T_f$.
In the experiments, $T_i=115$~K and $P_i\approx 0$, and $T_f\approx 205$~K.  

In our simulations, we chose $T_i=80$~K, the temperature at which the compression/decompression cycle to produce the samples of \hdad have been conducted in this and in several previous studies of ST2.
We note that the density of HDA samples recovered at ambient pressure varies~\cite{LGreview}.  In order to test if variations in $\rho_i$ influence the results of the heating-decompression process studied here, we recover starting samples of \hdad at $T_i=80$~K at a set of four different initial densities $\rho_i \in \{1.30,~1.36,~1.43,~1.48\}$~g/cm$^3$.  
Due to the variations that occur across the 10 independent decompression runs used to produce the recovered \mbox{HDA-d} samples, the pressures for the samples at each density varies over a finite interval.
The intervals of $P_i$ corresponding to the set of values of $\rho_i$ are  
$P_i \in \{(-200,-100),(-25,100),(375,550),(900,1150)\}$~MPa.
The range of samples so produced allows us to test if varying the density of the starting HDA-d sample affects the conditions at which HDA transforms to LDA during the decompression step of the heating-decompression process.

In addition, in order to provide a more general understanding of the thermodynamic pathway utilized in the experiments of Ref.~\cite{katrinNilsson}, we explore the effect of varying $T_f$, the temperature at the end of the heating stage and at which the isothermal decompression stage is carried out.  We vary $T_f$ over the 
range $100$ to $300$~K, which encompasses the glassy state 
(approximately $T<210$~K) and the liquid state (approximately $T>210$~K),
both below and above the temperature of the LLCP for ST2. 

For the isochoric heating stage of the heating-decompression process, the heating rate employed in our simulations is $q_T=30$~K/ns.  
This is approximately 100 times slower that the experimental heating rate.  While we could conduct simulations at the experimental rate, using $q_T=30$~K/ns allows us to compare our results to those obtained in previous ST2 simulation studies of amorphous ice that used the same rate.  This rate also allows our samples to pass through the glass transition temperature $T_g$ for the simulated system and access the equilibrium liquid state (when $T_f>T_g$) during the heating process itself.  Thus by tuning $T_f$, we are choosing whether the decompression stage starts with a glass (HDA) or a liquid (HDL).

For the isothermal decompression stage that follows the heating stage, 
the decompression rate in our simulations is $q_P=300$~MPa/ns.
This decompression rate is
approximately $100$ times faster than the rate estimated in the 
experiments~\cite{katrinNilsson}.  
A subset of our runs have been reproduced using a slower rate of $q_P=30$~MPa/ns. 
These results are presented in Sec.~I of the Supplementary Material (SM) and do not 
qualitatively change our conclusions.  On this basis, we do not expect that a 
further reduction of the rate to the experimental value of $q_T=4.3$~MPa/ns will markedly 
affect our results.  

\begin{figure*}      
\centerline{
\includegraphics[width=8cm]{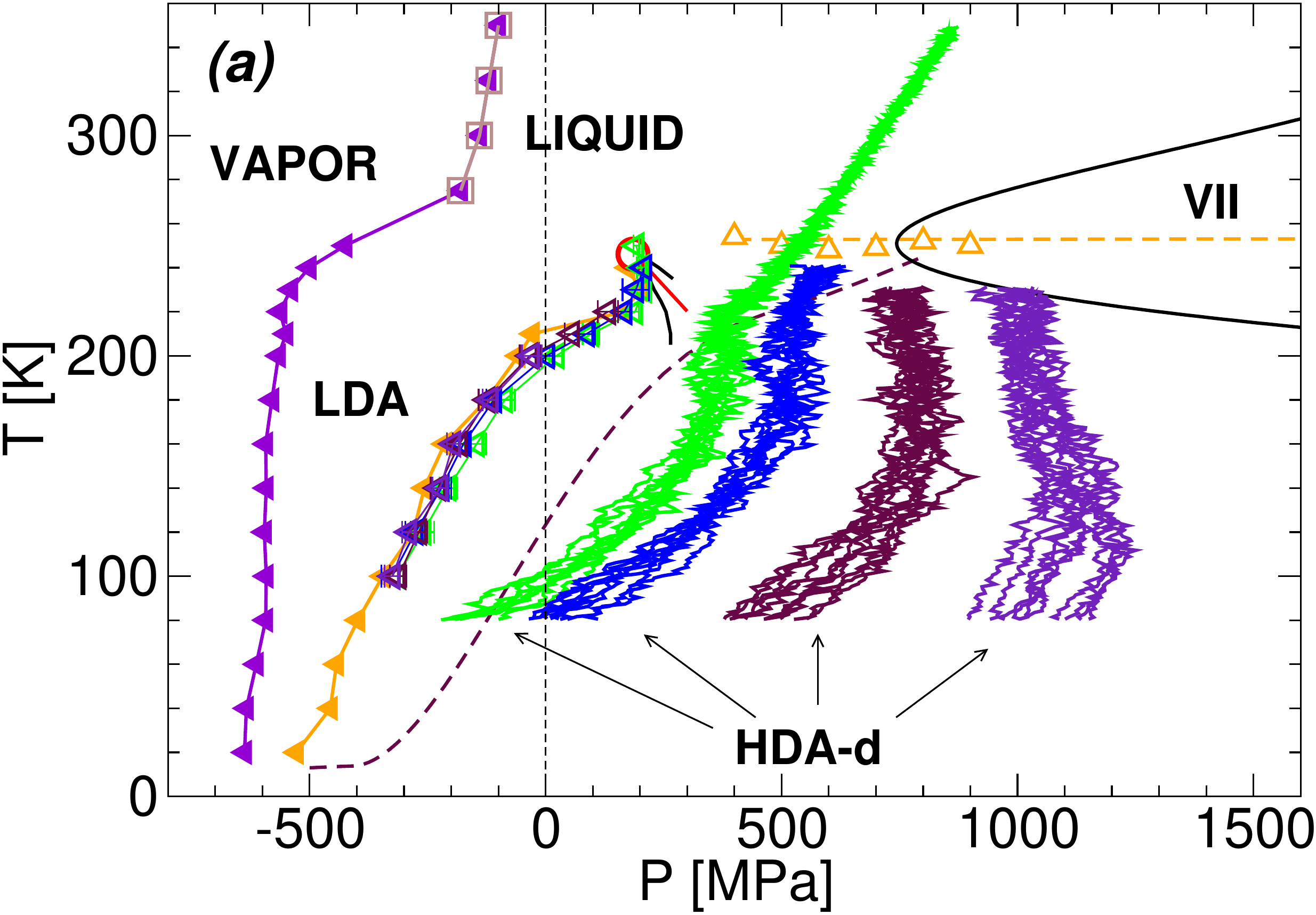}
\includegraphics[width=8cm]{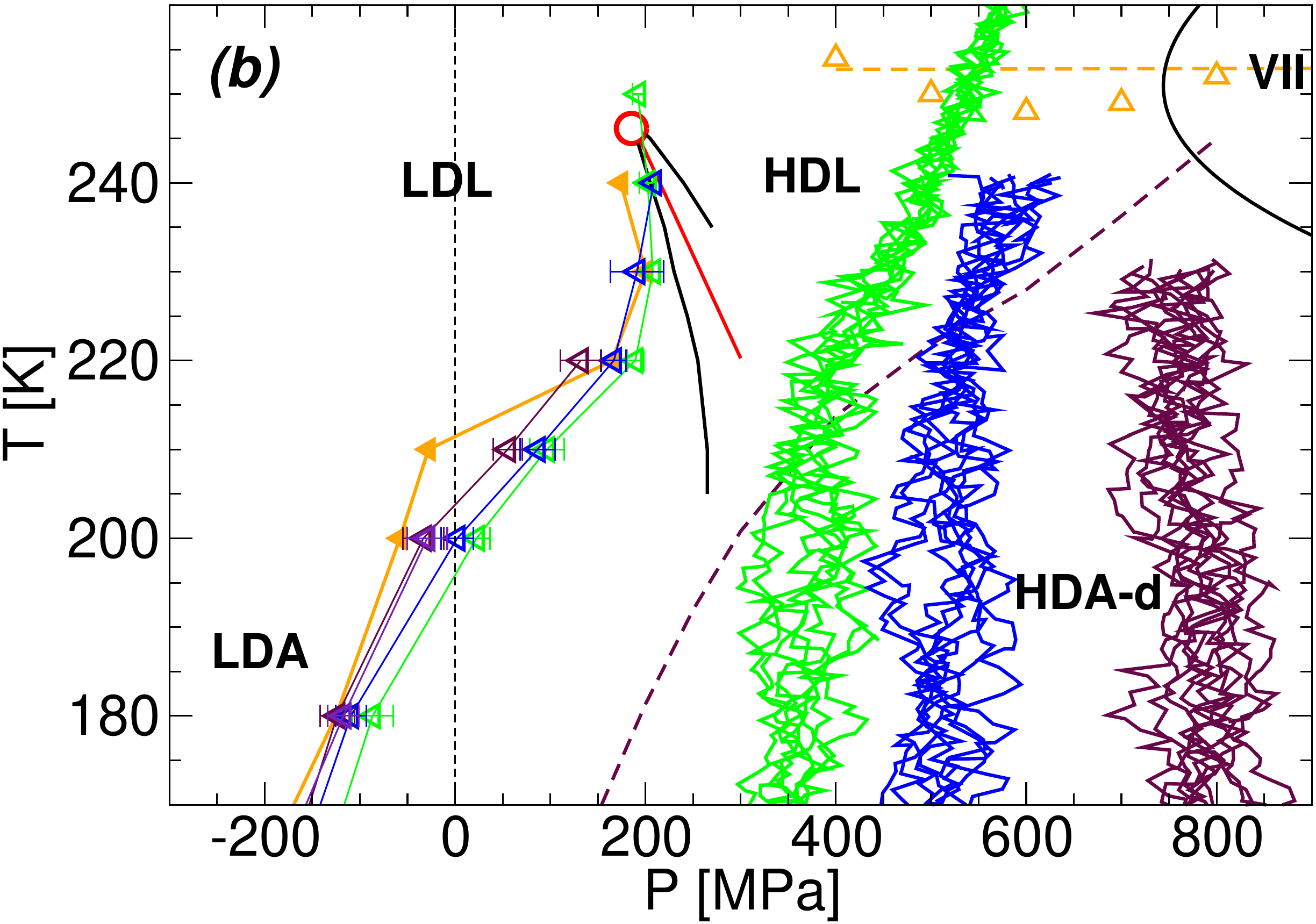}
}
\caption{(a) Temperature as function of pressure upon
heating HDA-d at $\rho_i=1.30$ (green lines), $\rho_i=1.36$ (blue lines),
 $\rho_i=1.43$ (maroon lines), and $\rho_i=1.48$~g/cm$^3$ (indigo lines).
The starting HDA-d configurations are obtained by decompressing HDA at $T=80$~K
from $P=1700$~MPa until the system reaches the target density $\rho_i$;
see red lines in Fig.~\ref{HDALDALIQ-Rates} for the case $q_P=300$~MPa/ns.
Heating trajectories are shown only for temperatures at which
crystallization is absent.  
Empty left-triangles represent the HDA-to-LDA transformation obtained
upon isothermal decompression of the HDA-d samples heated at $\rho_i$;
see Fig.~\ref{AllHDAd-Decomp}.
For comparison, also included are the HDA-to-liquid (at high pressure) and HDA-to-LDA (at low and negative pressure) transformations
upon isobaric heating (maroon dashed-line) from Ref.~\cite{chiu2}. 
Crystallization to ice VII occurs upon isothermal compression at $P>700$~MPa 
and $T=210-300$~K (solid black line)~\cite{chiu1}, and during 
isobaric heating of HDA at $T\approx 250$~K and $P>400$~K
 (orange up-triangles and dashed line)~\cite{chiu2}.
Solid orange and indigo left-triangles represent, respectively, the 
HDA/HDL-to-LDA/LDL and LDA/LDL-to-vapor transformations
during isothermal decompression of HDA samples 
with no heating treatment, reported in Ref.~\cite{chiu1}. 
Squares locate the equilibrium liquid-to-vapor spinodal line.
(b) Magnification of panel (a).
All compressions/decompressions are performed at rate $q_P=300$~MPa/ns;
all isochoric/isobaric heatings are performed at the rate $q_T=30$~K/ns. 
}
\label{AllHDAd}
\end{figure*}

\begin{figure*}      
\centerline{
\includegraphics[width=7cm]{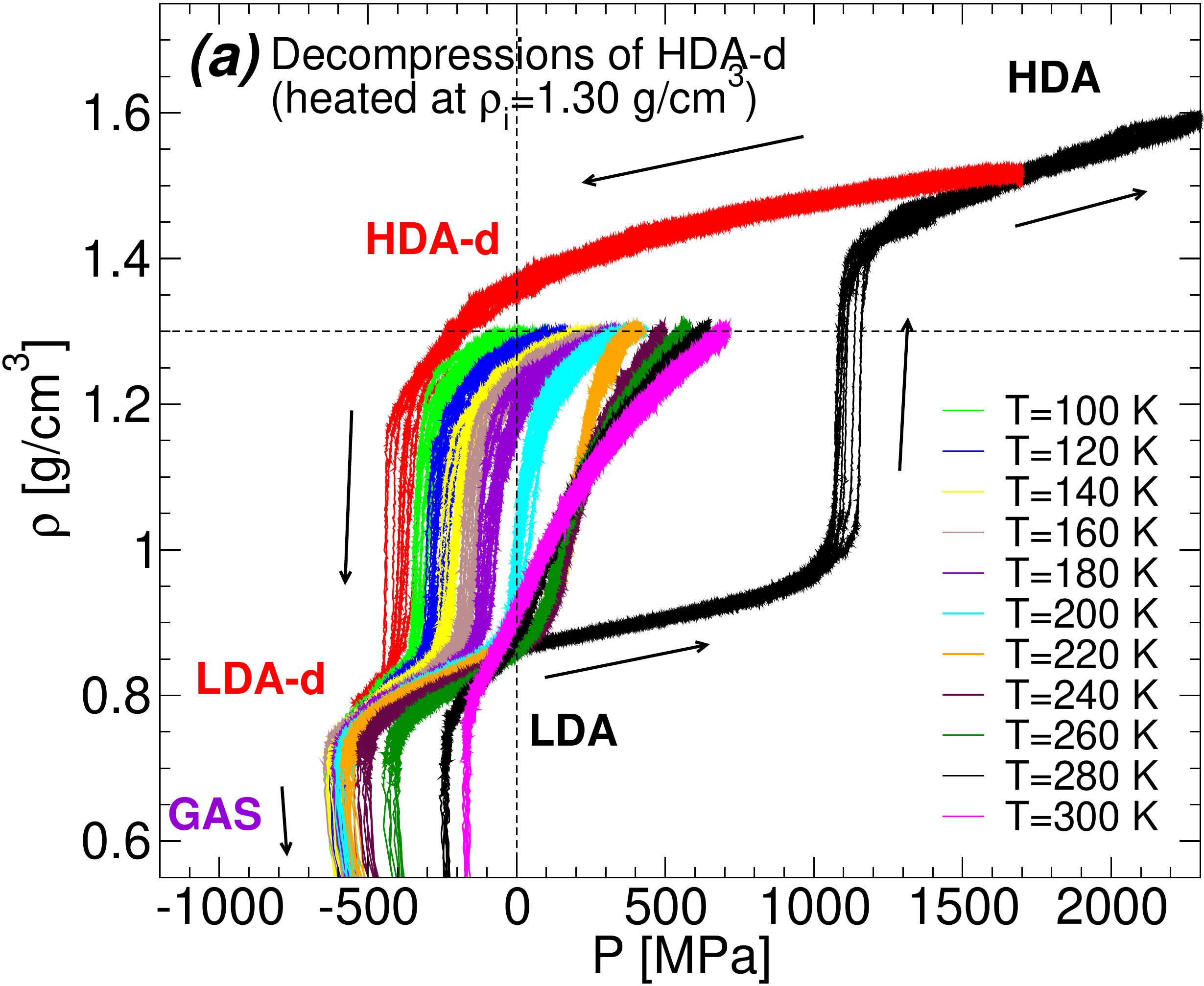}
\includegraphics[width=7cm]{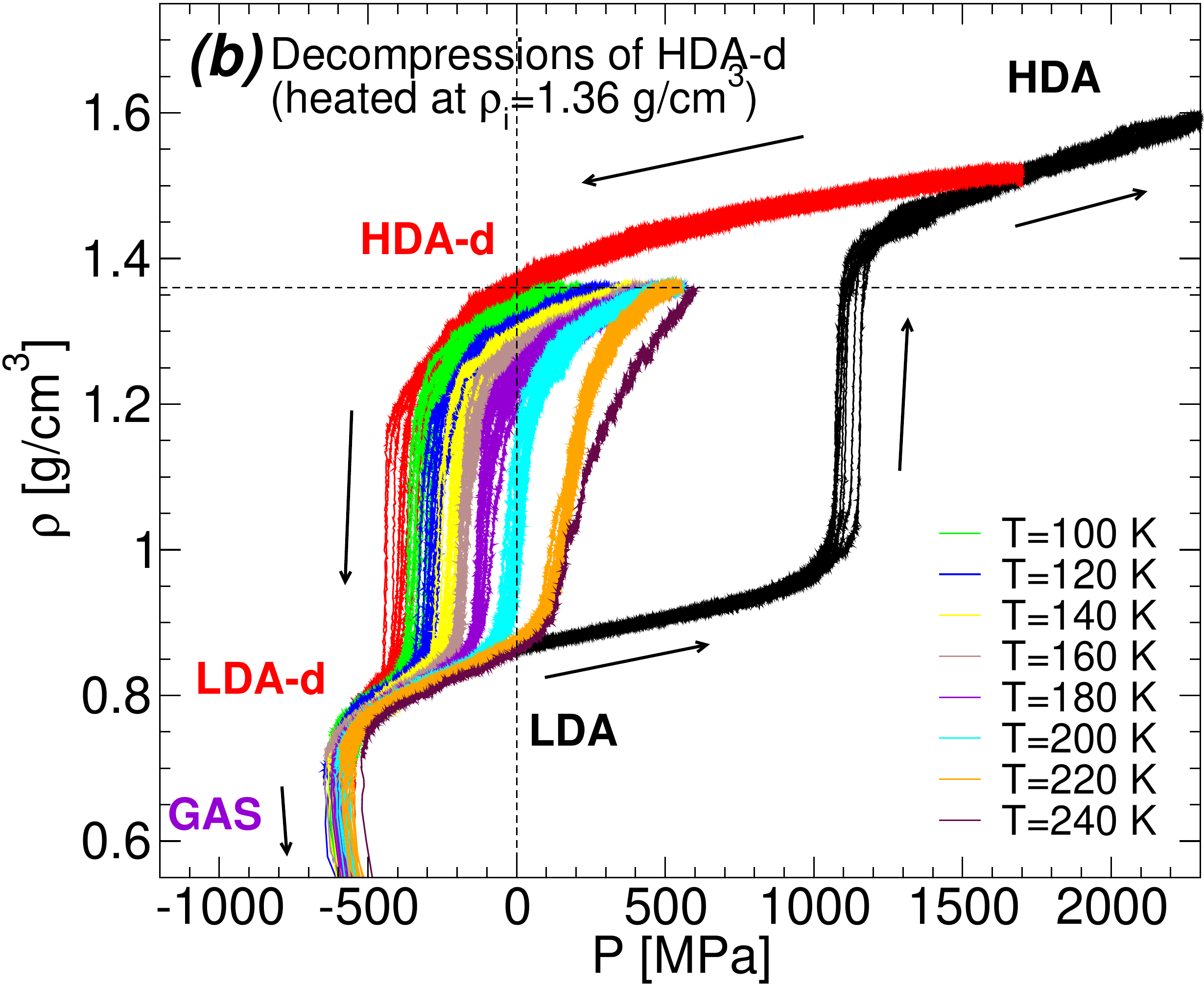}
}
\centerline{
\includegraphics[width=7cm]{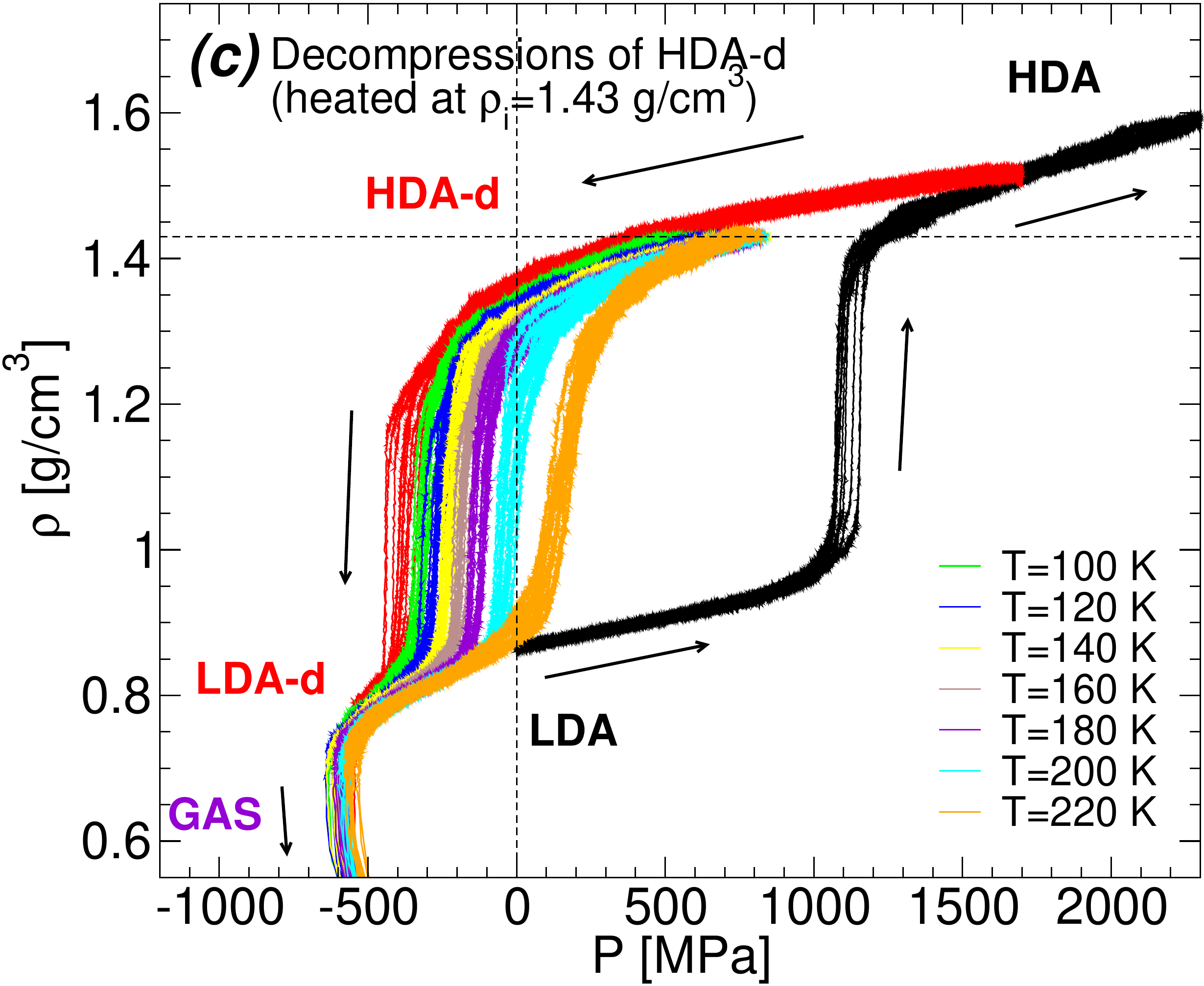}
\includegraphics[width=7cm]{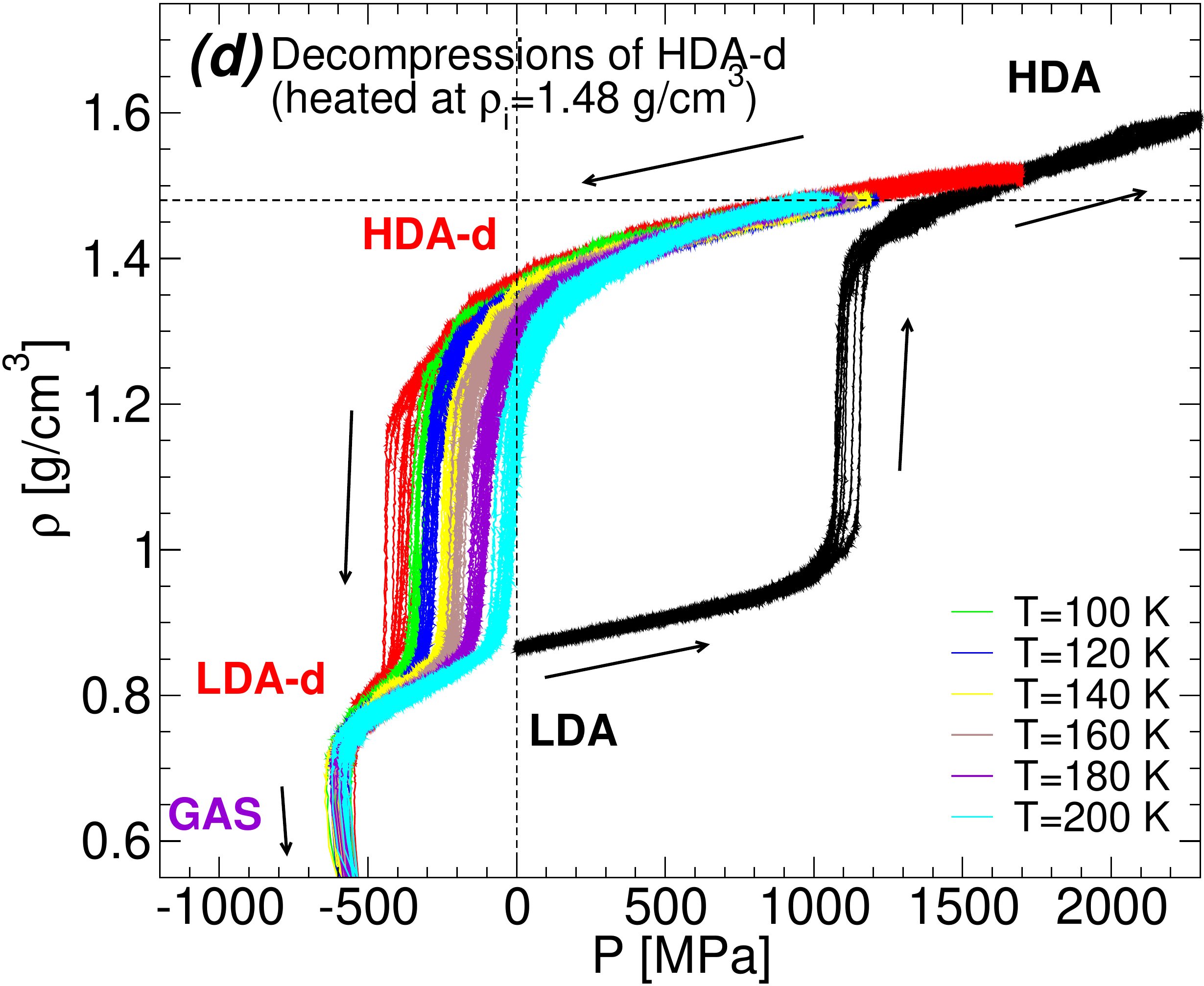}
}
\caption{Density as function of pressure during the isothermal 
decompression of HDA-d 
samples at different temperatures $T$. The HDA-d samples are prepared 
by isochoric heating at density $\rho_i$, from $T=80$~K to 
the target temperature $T$.
(a) $\rho_i=1.30$~g/cm$^3$, 
(b) $\rho_i=1.36$~g/cm$^3$, 
(c) $\rho_i=1.43$~g/cm$^3$, and 
(d) $\rho_i=1.30$~g/cm$^3$.
Also included are the pressure-induced LDA-to-HDA and HDA-to-LDA
transformations at $T=80$~K 
from Fig.~\ref{HDALDALIQ-Rates} (black and red lines).  
All compression/decompressions are performed at $q_P=300$~MPa/ns.
Some HDA-d samples crystallize during decompression at 
approx $T>230$~K and are not included for clarity.
}
\label{AllHDAd-Decomp}
\end{figure*}

\section{Results}
\label{results}

\subsection{Isochoric heating}


The schematic diagram shown in Fig.~\ref{schemeProcess} assumes that 
during the isochoric heating process, 
the pressure of HDA increases with increasing 
temperature (red arrow).  However, the experiments of Ref.~\cite{katrinNilsson} 
do not provide 
the pressure of the system upon heating or decompression.  
We show in Fig.~\ref{AllHDAd} 
the pressure as directly measured in our simulations during the isochoric heating of \hdad 
samples with different values of $\rho_i$, which therefore begin the heating process at different values of $P_i$.
We find that for our smallest value of $\rho_i=1.30$~g/cm$^3$, corresponding to samples with the lowest $P_i$, the pressure of the HDA-d samples indeed increases upon heating, confirming the assumption of  Fig.~\ref{schemeProcess} and Ref.~\cite{katrinNilsson}.  However, the variation of $P$ with $T$ is not linear, or even monotonic, up to $T_f\approx 220$~K at the heating rate used here.
For larger $\rho_i$, $P$ initially increases with $T$, and then reaches a maximum and decreases with $T$ as $T_f\to 220$~K.
 {\color{black}  We show in the SM (Sec.~II) that at the much faster heating rate of $q_T=3000$~K/ns, which is closer to the experimental rate, the variation of $P$ with $T$ for $\rho_i=1.30$~g/cm$^3$  approaches a linear relation~\cite{katrinNilsson}.  This result is consistent with the expectation that in the limit of infinitely fast heating, the increase in $P$ with $T$ will be dominated by the ideal gas contribution to $P$, which is linear in $T$ at fixed density.}

To place these results in context, we also show in Fig.~\ref{AllHDAd} 
three boundaries obtained in earlier work characterizing the behavior of the ST2 system.
(i)~We show the boundary in the $P$-$T$ plane 
where crystallization to the high-pressure ice VII phase was observed 
in Ref.~\cite{chiu1} during isothermal compression of ST2 water ($q_P=300$~MPa).
This crystallization region occurs inside the nose-shaped area indicated
by the black solid line at $P>700$~MPa and centered at $T\approx 250$~K.   
(ii)~The orange up-triangles at $T=250$~K indicate the 
temperature at which HDA/HDL crystallizes during isobaric heating (with $q_T=30$~K/ns) for
 $P>400$~MPa. 
(iii)~We also show the boundary in the $P$-$T$ plane at which 
the HDA-to-LDA and HDA-to-HDL transformations occur upon 
isobaric heating, as reported in Ref.~\cite{chiu2}.  
As shown in Refs.~\cite{MySciRep,chiu2}, these HDA-to-LDA and HDA-to-HDL
 transformations define a single line
in the $P$-$T$ plane that represents the limit of stability of HDA relative to LDA or HDL
during isobaric heating. This metastability limit is indicated 
approximately by the maroon dashed line in Fig.~\ref{AllHDAd}
(for the case $q_T=30$~K/ns).  
We note that the HDA-to-LDA transformation line in the $P$-$T$ plane, 
when measured using ultrafast rates, is sensitive to the procedure followed.  
For example, in Fig.~\ref{AllHDAd} the HDA-to-LDA transformation induced by 
isobaric heating with $q_T=30$~K/ns at low and negative pressures (dashed maroon line) differs 
from the HDA-to-LDA transformation induced by isothermal decompression with $q_P=300$~MPa/ns (orange left-triangles).
As expected, these differences in the locations of the HDA-to-LDA transformation lines in the $P$-$T$ plane decrease as the rates employed ($q_T$ or $q_P$) decrease~\cite{chiu1,chiu2}. 

We observe crystallization 
during our isochoric heating runs at all densities $\rho_i$.
As shown in Sec.~III of the SM, crystallization is readily detected by monitoring the 
oxygen-oxygen radial distribution function of the system upon heating, and occurs in the range $T=230$-$260$~K.
At $\rho_i=1.30$~g/cm$^3$, 
2 out of the 10 independent runs crystallize.
At $\rho_i=1.36$~g/cm$^3$, 
9 out of the 10 independent runs crystallize.
For all larger values of $\rho_i$, all runs crystallize.
In Fig.~\ref{AllHDAd}, we plot the pressure for each heating run only up to the temperature where crystallization occurs.
The crystallization behavior observed here is largely consistent with the boundaries for the onset of crystallization found in previous work during isothermal compression (solid black line) and isobaric heating (orange up-triangles).  The exceptional case occurs for $\rho_i=1.30$~g/cm$^3$, where 8 out of 10 runs resist crystallization, allowing the system to be heated well into the liquid phase up to $350$~K.  The absence of crystallization in an isochoric run which passes through a state point at which an isobaric run would reliably crystallize is likely a finite-size effect in our simulations.

We also find that all our runs at $\rho_i=1.30$ and $1.36$~g/cm$^3$ crossover from the glassy HDA state to the liquid HDL state in the range $T=210$ to $220$~K, regardless of whether the runs crystallize at higher $T$.  This crossover corresponds to the glass transition temperature $T_g$ encountered when heating the ST2 system at these densities at $q_T=30$~K/ns.  As shown in SM Sec.~III, at $T_g$ the mean-square displacement of the molecules becomes non-negligible, and the temperature dependence of the pressure and total energy shows the characteristic change in slope consistent with passing through the glass transition.  Similarly, in Fig.~\ref{AllHDAd}, we see that the variation of $P$ with $T$ at $\rho_i=1.30$~g/cm$^3$ changes behavior and becomes linear for $T>210$~K, for those runs that do not crystallize.   This behavior is consistent with the $T_g$ boundary (maroon dashed line) found in previous work from isobaric heating runs at the same $q_T=30$~K/ns.  That is, our runs at $\rho_i=1.30$ and $1.36$~g/cm$^3$ enter a window of $T$ between $T_g$ and the crystallization temperature within which the HDL phase is accessed in our simulations.  
As shown in Fig.~\ref{AllHDAd},
our runs at $\rho_i=1.43$ and $1.48$~g/cm$^3$ crystallize before they reach $T_g$, and so the HDL phase is not observed in these runs prior to crystallization.

Based on the above observations, our simulations thus confirm that rapid isochoric heating provides a method to drive a sample of HDA at low-$T$ and ambient-$P$ to a location in the phase diagram corresponding to the equilibrium HDL phase at pressures above the coexistence line of the LLPT, just as proposed in the experiments of Ref.~\cite{katrinNilsson}.

\subsection{Isothermal decompression}


We next carry out isothermal decompression at fixed $T=T_f$ of HDA-d samples heated at constant $\rho_i$ to various temperatures $T_f$ in the range from $100$ to $300$~K.
Fig.~\ref{AllHDAd-Decomp}(a) shows $\rho$ for each sample as function of $P$
during these isothermal decompression runs for different $T$.  All these samples have an initial density,
 prior to decompression, of $\rho_i=1.30$~g/cm$^3$ and, depending on $T$,
they correspond to HDA (for $T_f < 210$~K) or HDL (for $T_f >210$~K).  
{\color{black} As shown in Fig.~\ref{AllHDAd-Decomp}(a), all samples exhibit a density 
change during decompression characteristic of the transformation to an LDA or LDL-like state.
The sharpness of the density change depends on $T$.
In particular, for $T<200$~K where the system is
a glass at all $P$, the HDA-to-LDA transformations are reminiscent of a sharp,
discontinuous first-order phase transition.  
Similar abrupt density changes occur during the pressure-induced LDA-HDA transformation
cycles at $T=80$~K~\cite{chiu1} (red and black lines).
At $T>200$~K, the high-density to low-density transformation as a function of $P$ becomes smoother but still follows a curve where $\rho$ has a well-defined inflection point.
At the highest $T>280$~K this inflection point is weak or absent, consistent with entering the $T$-range well above the LLCP where such an inflection would also be weak or absent in the equilibrium system; see Fig.~\ref{schemeProcess}.}  
The pattern of behavior shown in Fig.~\ref{AllHDAd-Decomp}(a) is observed for all values of $\rho_i$ studied here, as shown in Figs.~\ref{AllHDAd-Decomp}(b)-(d), demonstrating that the occurrence of the high-density to low-density transformation is insensitive to the density at which HDA or HDL is prepared.

In Sec.~IV of the SM, we test if changing the preparation pathway for our starting HDA samples influences our results.  Specifically, we prepare HDA samples by instantaneous isochoric quenches from the equilibrium liquid state to $T=80$~K.  After adjusting the density of these quenched samples to match $\rho_i=1.30$~g/cm$^3$, they are subjected to the same two-stage process of isochoric heating and isothermal decompression applied to our HDA-d samples.  We find that 
the high-density to low-density transformation is again observed, demonstrating 
that our results are independent of the method followed 
to prepare the starting HDA sample of the heating runs.

Notably, the pressure of the high-density to low-density transformation 
at fixed $T$ is independent of the
 density $\rho_i$ at which HDA-d is prepared.
This is shown in Fig.~\ref{AllHDAd}, 
where the pressure of this 
transformation $P_{\rm H\to L}$ is presented as a function of $T$ for each of the values of $\rho_i$ studied here.
We define $P_{\rm H\to L}$ as the pressure, averaged 
 over the $10$ independent runs, at which the sample density passes through $\rho=1.05$~g/cm$^3$.  As shown in Fig.~\ref{AllHDAd-Decomp}(a)-(d),
 this density corresponds approximately to the midpoint 
of the high-density to low-density transition.   
Also shown in Fig.~\ref{AllHDAd}, all values of $P_{\rm H\to L}$ coincide with the 
$P_{\rm H\to L}$ values obtained in Ref.~\cite{chiu1} via decompression of
HDA samples prepared by compression of LDA at fixed $T$.
This demonstrates that the behavior observed during decompression is robustly insensitive to the method by which HDA samples under pressure are prepared, whether it be by ultrafast isochoric heating (as studied here and in Ref.~\cite{katrinNilsson}) or by isothermal compression~\cite{chiu1}.

Our results also show that regardless of whether the isothermal decompression process starts with HDA or HDL, the signature of the LLPT is observed, in the form of an inflection point in the variation with $P$ of the order parameter $\rho$.  Our results are thus consistent with the interpretation of the evolution of the structure factor observed in Ref.~\cite{katrinNilsson} during the decompression stage of the experiment.  Regardless of the initial state of the sample prior to decompression in our simulations, we note that at the decompression rate used here, none of the simulations that we conduct for $T<T_c$ access the equilibrium LDL state upon decompression; the low-density states in our simulations for $T<T_c$ are always
 glassy LDA or non-equilibrium LDL.  In the experiments of Ref.~\cite{katrinNilsson}, 
the decompression rate is approximately $100$ times slower, and in addition, crystallization does not begin 
until approximately $10~\mu$s after the end of the IR heating stage.  Ref.~\cite{katrinNilsson} shows 
that this time scale is sufficient to access equilibrium liquid LDL at the end of the 
decompression stage in the experiment (at $P \approx 0$).

\section{discussion}

{\color{black} In summary, the results of our simulations as shown in Figs.~\ref{AllHDAd} and \ref{AllHDAd-Decomp}
are consistent with the experiments of Ref.~\cite{katrinNilsson} 
and support the interpretation of the thermodynamic pathway presented there.  We have shown that a process of rapid isochoric heating applied to HDA at $P=0$ can be used to access the HDL phase at pressures above the coexistence line of the LLPT.  Furthermore, when such a sample is decompressed isothermally, it will pass through the conditions of the LLPT and convert to the low-density phase, which is LDA or non-equilibrium LDL in our simulations and which is equilibrium LDL in the experiments.
Furthermore, depending on the decompression temperature,
the high-density to low-density transformation observed in our simulations can be smooth or sharp, but so long as $T<T_c$, the transition observed in our rapid decompression runs exhibits an inflection of the density as a function the pressure consistent with the discontinuous HDL-to-LDL transition of the equilibrium system.
That is, despite the use of very fast heating and decompression rates, the influence of the underlying equilibrium behavior of the system remains observable during the strongly out-of-equilibrium process studied here.

Importantly, and as anticipated in Ref.~\cite{katrinNilsson}, 
we find that due to the ultrafast decompression rate, neither the experimental nor the computational high-density to low-density transformation at $P_{\rm H\to L}$ occurs at the equilibrium LDL-HDL coexistence pressure $P_{\rm coex}$. 
Rather, as shown in Fig.~\ref{AllHDAd}, for $T < T_c$ we find that $P_{\rm H\to L} < P_{\rm coex}$. The position of the $P_{\rm H\to L}$ curve depends on several factors, especially the decompression rate, and so it may vary widely between simulations and experiments.  Nonetheless, Fig.~\ref{AllHDAd} shows that the $P_{\rm H\to L}$ curve observed in the ultrafast process studied here places a lower bound on the location of the equilibrium coexistence line of the LLPT.
Notably, the $P_{\rm H\to L}$ curve also converges to the location of the equilibrium LLCP in 
the $P$-$T$ plane as $T \rightarrow T_c$.  
Fig.~\ref{AllHDAd} further shows that the $P_{\rm H\to L}$ curve overlaps with the HDL-to-LDL spinodal line evaluated from ST2 simulations of the equilibrium liquid in the range $235~{\rm K} < T< T_c$~\cite{poole2005}. 
 As shown in SM Sec.~I, we find that the range of $T$ over which this
 overlap occurs increases when the decompression rate decreases.
Our work thus demonstrates that if the pressure corresponding to the $P_{\rm H\to L}$ curve as a function of $T$ could be measured in an experiment, it would be possible to locate the equilibrium LLCP in real water using 
the ultrafast process pioneered in Ref.~\cite{katrinNilsson}.

A broader conclusion may also be drawn by combining the present results with those reported in 
Refs.~\cite{MySciRep,chiu1,chiu2,ourST2PEL-1,ourST2PEL-2,ourST2PEL-3}.  These studies collectively document the behavior of glassy ST2 water when subjected to a wide variety of heating, cooling, compression and decompression processes.  As highlighted here, so long as the relevant rates $q_T$ and $q_P$ are constant, important boundaries defining the response of the system (e.g. $T_g$, $P_{\rm H\to L}$, crystallization) can be predicted for one process (e.g. isochoric heating) using results found from another process (e.g. isothermal compression).  That is, for practical purposes, one can define a phase diagram
for glassy and supercooled water with well-defined boundaries between the LDA, HDA, LDL, and HDL when $q_T$ and $q_P$ are fixed.  It would be interesting to explore the implications of this finding for the definition of an out-of-equilibrium free energy function $F(N,V,T, q_T,q_P)$ for glassy water, and the corresponding phase diagram, 
that explicitly includes the rates as ``state variables".

\section*{Supplemental Material}

In Sec.~I of the SM, we test whether our results are sensitive to the 
compression/decompression rates employed.  
In Sec.~II, we show how increasing the heating rate during isochoric heating affects the temperature dependence of the pressure and energy.
Sec.~III provides a brief description of
dynamical and structural properties used to identify the HDA-to-HDL transformation 
and crystallization during the isochoric heating of HDA.
In Sec.~IV, we show that the results presented in the main manuscript 
are independent of the process followed to
prepare the recovered HDA samples at $T=80$~K and pressure $P_i$.

\begin{acknowledgments}
PHP acknowledges the support of the Natural Sciences and Engineering Research Council of Canada (NSERC), Grant Number RGPIN-2017-04512.  We also thank ACENET and Compute Canada for support.
We thank K. Amann-Winkel, K. H. Kim, and A. Nilsson for valuable discussions.
We dedicate this work to the memory of C. Austen Angell, a dear friend and great experimentalist, who was also an innovator in the use of computer simulations to illuminate experimental results for water and many other systems.\end{acknowledgments}

\section*{Data Availability Statement}

The data that support the findings of this study are available from the corresponding author upon reasonable request.








\clearpage

\onecolumngrid

\centerline{\bf {Supplemental Material:}}

\centerline{\bf {Liquid-liquid phase transition in 
simulations of ultrafast heating}}
\centerline{\bf {and decompression of amorphous ice}}

\bigskip
\centerline{Nicolas Giovambattista$^{1,2}$ and Peter H. Poole$^{3}$}
\smallskip
{\it{
\centerline{$^1$Department of Physics, Brooklyn College of the City
   University of New York, Brooklyn, NY 11210, United States}
   
\centerline{$^2$Ph.D. Programs in Chemistry and Physics,
The Graduate Center of the City University of New York,} 
\centerline{New York, NY 10016, United States}
\centerline{$^3$Department of Physics, St. Francis Xavier University,
     Antigonish, Nova Scotia B2G 2W5, Canada}
}}

\centerline{(Dated: {\today})}
\bigskip

\setcounter{equation}{0}
\setcounter{figure}{0}
\setcounter{section}{0}
\setcounter{table}{0}
\setcounter{page}{1}
\makeatletter

\renewcommand{\theequation}{S\arabic{equation}}
\renewcommand{\thefigure}{S\arabic{figure}}
\renewcommand{\thesection}{\Roman{section}}

\twocolumngrid

\section{Simulations Using a Slower Compression/Decompression Rate}
\label{rateSec}

In the main manuscript, the HDA forms recovered at $T=80$~K and pressure $P_i$,
which are used as starting configurations
in the isobaric heating runs, correspond to HDA-d. In particular, 
all compression/decompression runs are performed at the rate $q_P=300$~MPa/ns.
In this section, we repeat the heating-decompression simulations 
described in the main manuscript but using a slower compression/decompression rate of $q_P=30$~MPa/ns, 
closer to the estimated experimental rate. 

\begin{figure*}     
\centerline{
\includegraphics[width=8cm]{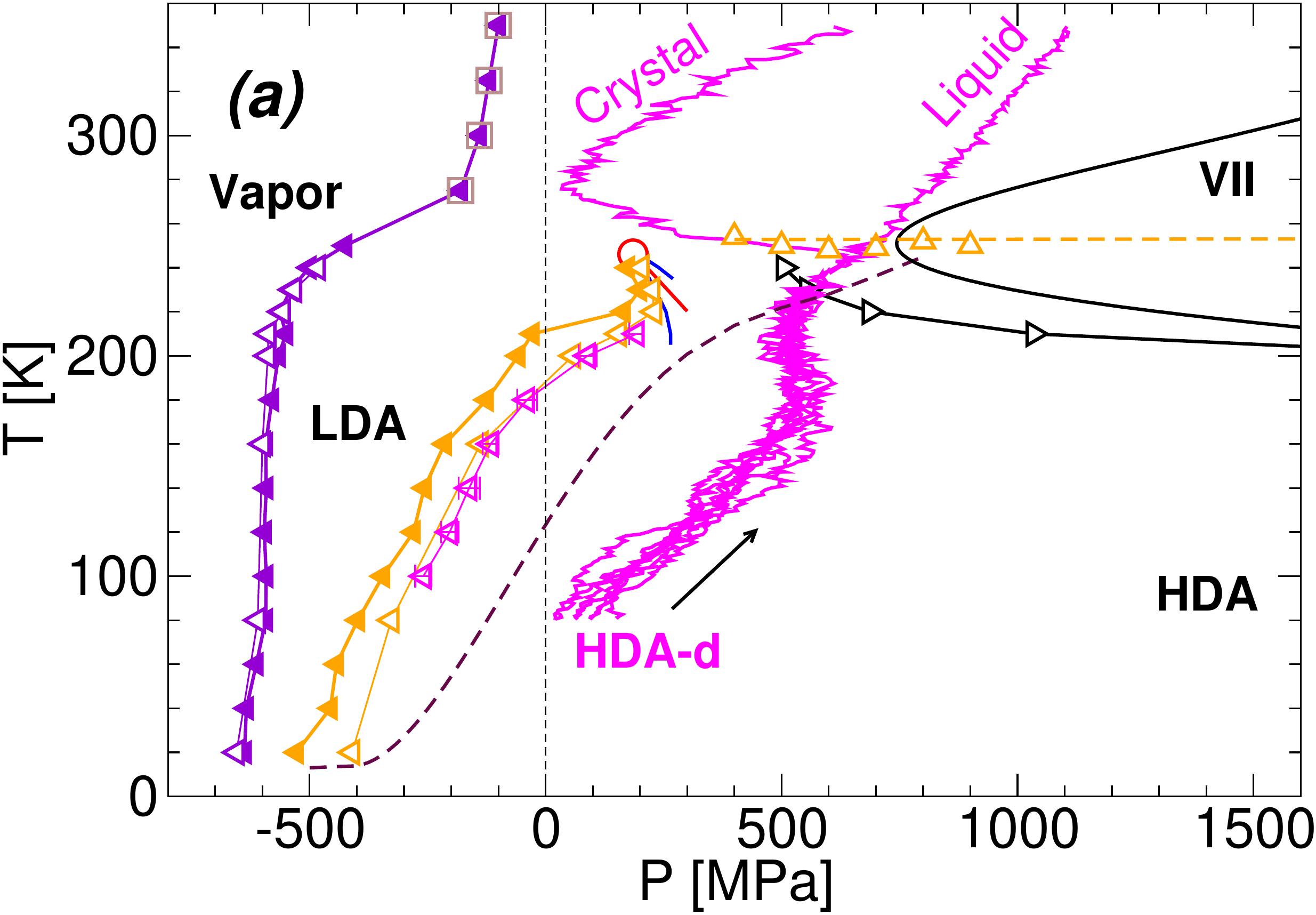}
\includegraphics[width=8cm]{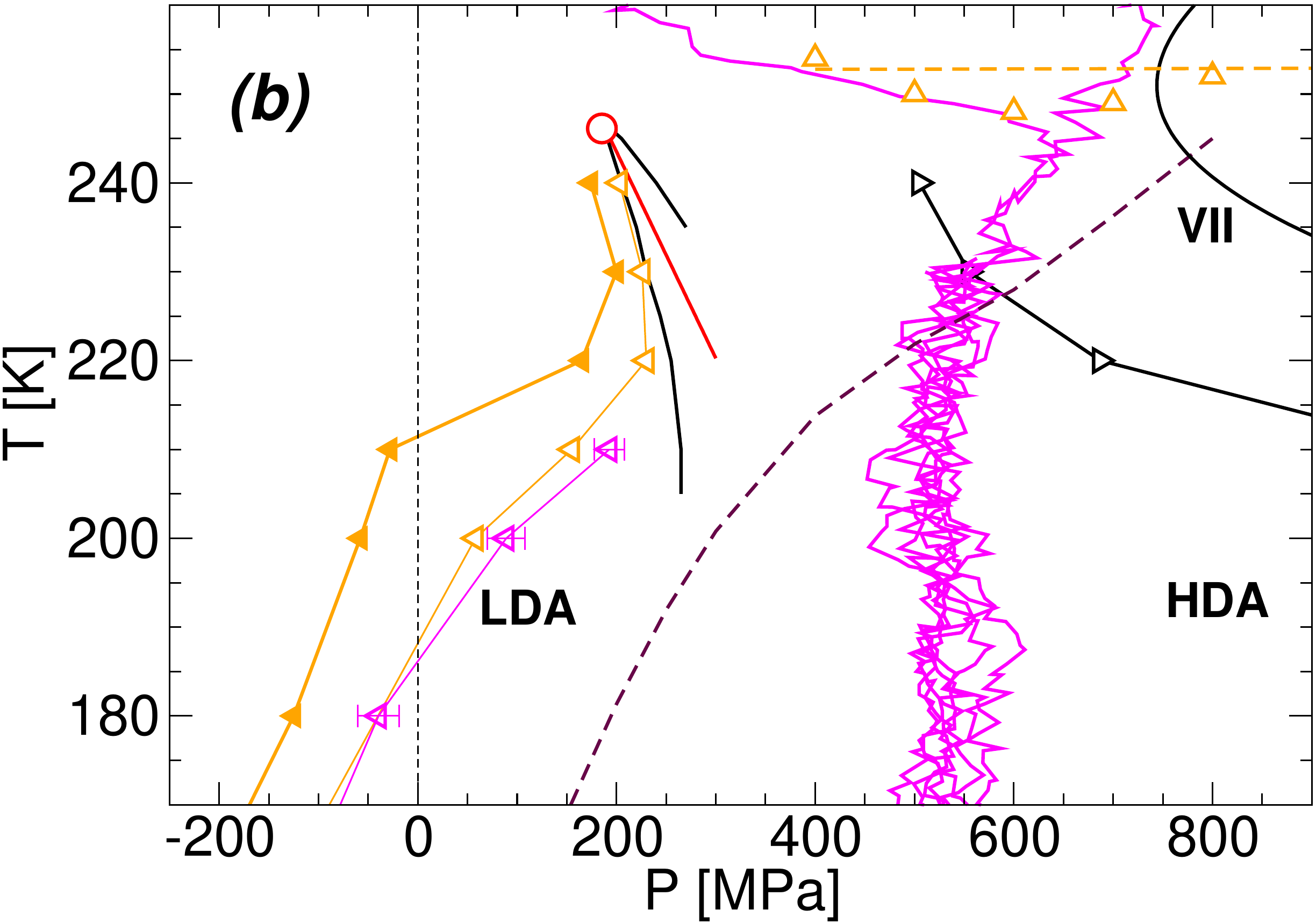}
}
\caption{Same as Fig.~3 of the main manuscript for the case of HDA-d samples obtained using
a slower compression/decompression rate $q_P=30$~MPa/ns.
Temperature as function of pressure upon
heating HDA-d at $\rho_i=1.36$~g/cm$^3$ (magenta lines) with $q_T=30$~K/ns.
The starting HDA-d configurations are obtained by decompressing HDA at $T=80$~K
from $P=1500$~MPa until the system reaches the target density $\rho_i$; see green lines in Fig.~2b of the main manuscript for the case $q_P=30$~MPa/ns.
Empty magenta left-triangles represent the HDA/HDL-to-LDA/LDA transformation obtained
upon isothermal decompression of samples formed during heating of HDA-d at $\rho_i$;
see also Fig.~\ref{HDAdRates-Decomps}.
Solid violet and orange left-triangles represent, respectively,
the pressure at which LDA fractures and the HDA-to-LDA transformation 
upon decompression at the fast rate $q_P = 300$~MPa
from Fig.~3 of the main manuscript.  The corresponding transformation lines obtained 
at $q_P=30$~MPa/ns are indicated by empty
 violet and orange left-triangles~\cite{chiu1,chiu2}.
Squares are the equilibrium liquid-to-vapor spinodal line. The maroon dashed line indicates the transformation of 
HDA to LDA (at low pressure) and HDA to HDL (at high pressure) upon isobaric heating. The orange up-triangles 
signal the temperature at which HDA crystallizes upon isobaric heating~\cite{chiu2}.
}
\label{HDAdRates}
\end{figure*}

First we obtain new samples of HDA-d by compression of LDA at $T=80$~K 
and $q_P=30$~MPa/ns from $P=0.1$~MPa to $P=1500$~MPa. The resulting HDA is then decompressed at $T=80$~K to $P_i\approx 0$, yielding HDA-d. 
For simplicity, we only consider the case where of starting HDA-d samples have a density $\rho_i=1.36$~g/cm$^3$ 
at $T=80$~K, giving $P_i$ in the range from $0$ to $150$~MPa.

These new HDA-d samples are then heated isochorically at the same rate used in the main manuscript, i.e., $q_T=30$~K/ns.  Fig.~\ref{HDAdRates} shows the trajectories of our ten independent runs upon heating HDA-d at constant $\rho_i=1.36$~g/cm$^3$.  Fig.~\ref{HDAdRates} is analogous to Fig.~3 of the main manuscript, and the results in these two figures are qualitatively identical.
For the HDA-d samples prepared at $q_P=30$~MPa/ns, we find that nine of the ten independent runs crystallize
upon heating. In Fig.~\ref{HDAdRates}, we show the single trajectory that extends into the liquid state without signs of crystallization.
For comparison, also included is one of the trajectories that exhibit crystallization.  The remaining trajectories are shown up to the temperature where crystallization occurs
($T \leq 240$~K).

The same boundary lines shown in Fig.~3 of the main manuscript (for $q_P=300$~MPa)
are included in Fig.~\ref{HDAdRates}: 
(i) The black solid line indicates the crystallization (nose-shaped) region at high pressures  determined during isothermal compression in Ref.~\cite{chiu1}.
(ii) The orange up-triangles at $T=250$~K indicate the
temperature at which HDA/HDL crystallize during isobaric heating at $P>400$~MPa ($q_T=30$~K/ns).
(iii) The maroon dashed-line represents the $P$-$T$ conditions at which
the HDA-to-LDA and HDA-to-HDL transformations occur upon
isobaric heating~\cite{chiu2} with $q_T=30$~K/ns.
(iv) The solid orange left-triangles indicate the HDA/HDL-to-LDA/LDL transformation
reported during isothermal decompression in Ref.~\cite{chiu1}. 
(v) The solid indigo left-triangles indicate the pressures at which the recovered LDA forms fracture upon isothermal decompression.
The boundary lines (i)-(v) are valid when all the 
 compression/decompression runs are performed with rate $q_P=300$~MPa/ns. 
 Reducing the rate to $q_P=30$~MPa/ns
does not affect (v); see indigo empty left-triangles. However, it increases the size of the crystallization region (i) at
 high pressures; the crystallization region obtained using $q_P=30$~MPa/ns is indicated 
by the black empty right-triangles. Similarly, reducing $q_P$ shifts the boundary (iv) towards higher 
pressures; the HDA/HDL-to-LDA/LDL transformation 
line obtained during isothermal decompression at $q_P=30$~MPa/ns is indicated by empty orange
 left-triangles (see Ref.~\cite{chiu1}).
One of the main points of Fig.~\ref{HDAdRates} is that the temperature at which HDA-d 
reaches the liquid state, $P\approx 550$~MPa and $T\approx 210-220$~K (see Sec.~\ref{crystalSec}),
 occurs at the intersection of the magenta lines with the maroon dashed-line,
 i.e., at the glass transition temperature of HDA determined independently from
 isobaric heating runs at the same heating rate $q_T=30$~K/ns employed here; see Ref.~\cite{chiu2}.  
 
\begin{figure}      
\centerline{
\includegraphics[width=7cm]{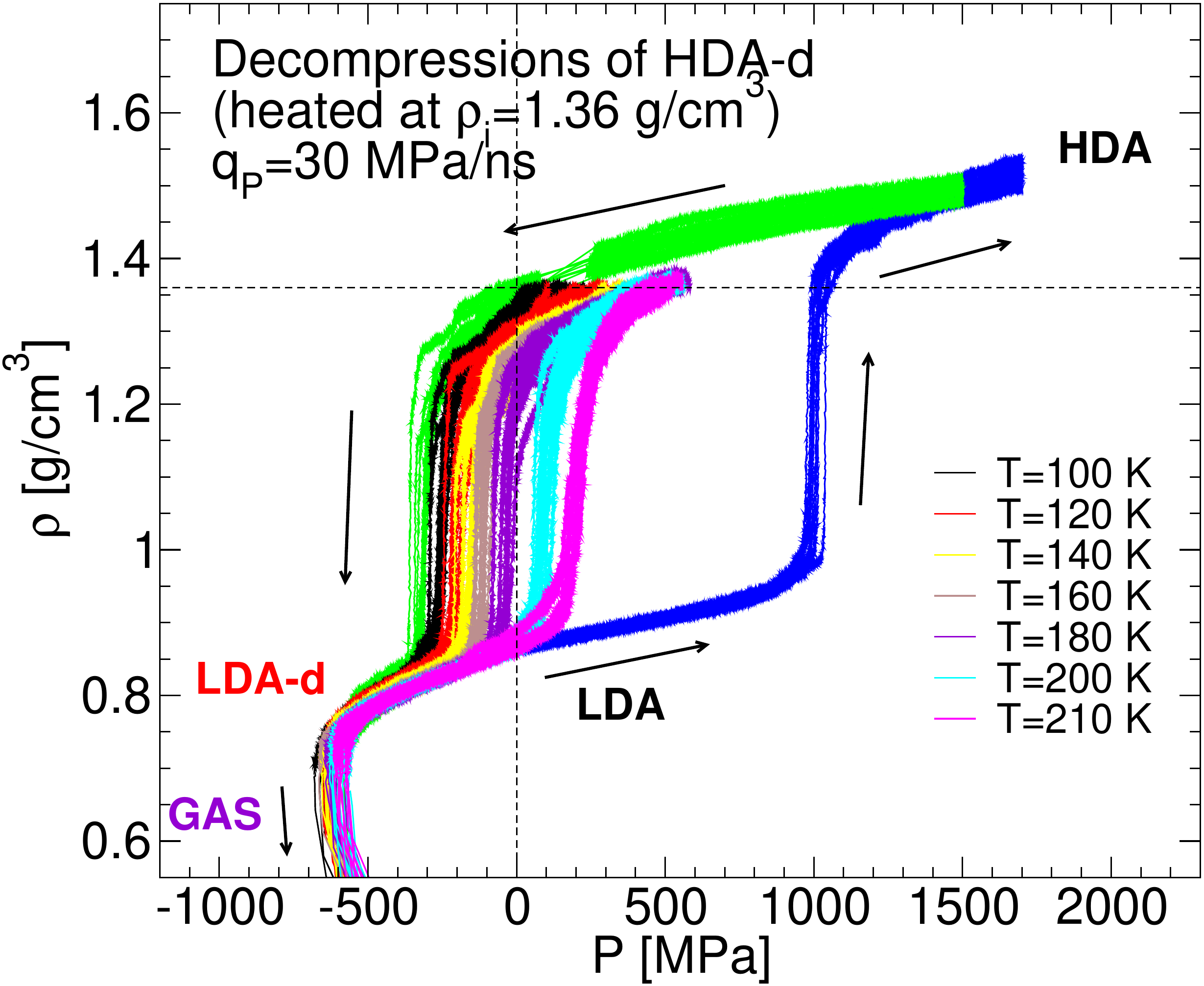}
}
\caption{Same as Fig.~4b of the main manuscript for the HDA-d samples obtained with
the slower compression/decompression rate $q_P=30$~MPa/ns.
 The starting HDA/HDL samples for the decompression runs at temperature $T$ 
 are prepared by isochoric heating of HDA-d at density $\rho_i=1.36$~g/cm$^3$, from $T=80$~K to
the target temperature $T$.
Blue and green lines are the pressure-induced LDA-to-HDA and HDA-to-LDA
transformations at $T=80$~K at the slower rate $q_P=30$~MPa/ns, as also shown in
 Fig.~2b of the main manuscript.
HDA-d samples crystallize during heating or decompression at $T>210$~K.
}
\label{HDAdRates-Decomps}
\end{figure}
 
The HDA/HDL samples obtained by isochoric heating to the temperature $T=T_f$ are then subjected to isothermal decompression at this $T$.
Fig.~\ref{HDAdRates-Decomps} shows the density as function of temperature 
for HDA/HDL samples decompressed at different $T \leq 210$~K. (At $T>210$~K crystallization occurs
 either during the heating runs or upon isothermal decompression). As reported in Fig.~4 of 
the main manuscript, we find that the HDA/HDL-to-LDA/LDL transformation is rather sharp
 at all $T \leq 210$~K. 
The HDA/HDL-to-LDA/LDL transformation pressures $P_{\rm H \to L}$ are included in
Fig.~\ref{HDAdRates} (magenta left-triangles). In agreement with the discussion in the 
main manuscript, $P_{\rm H \to L}$ (magenta left-triangles) overlaps
 with the HDA/HDL-to-LDA/LDL transformation pressures (empty orange left-triangles) 
obtained in Ref.~\cite{chiu1} upon isothermal decompression of HDA/HDL at $q_P=30$~MPa/ns.
That is, reducing $q_P$ shifts the decompression-induced HDA/HDL-to-LDA/LDL transformation
 towards higher pressures.
Nonetheless, the $P_{\rm H \to L}$ line determined during ultrafast heating-decompression simulations still coincides with the HDA/HDL-to-LDA/LDL transformation pressures determined during isothermal compression-decompression cycles of LDA to HDA and back, when conducted at the same rate $q_P$. 
 
\section{Increasing the rate of isochoric heating}
\label{fastSec}

Here we test the effect of increasing the rate of heating $q_T$ used in our isochoric heating runs that start with HDA-d at $T=80$~K and $\rho_i = 1.30$~g/cm$^3$.
Fig.~\ref{fast-heating} shows $P$ and $E$ as a function of $T$ using $q_T=30$, $300$ and $3000$~K/ns.  We find that as $q_T$ increases, the variation of $P$ with $T$ at fixed density becomes more linear, the result that would be expected when the ideal gas contribution to the pressure dominates.  Despite the very fast heating rates explored here, all our runs give similar values for $P$ and $E$ when the system accesses the liquid state for $T>T_g\approx 220$~K.

\begin{figure}      
\centerline{
\includegraphics[width=7cm]{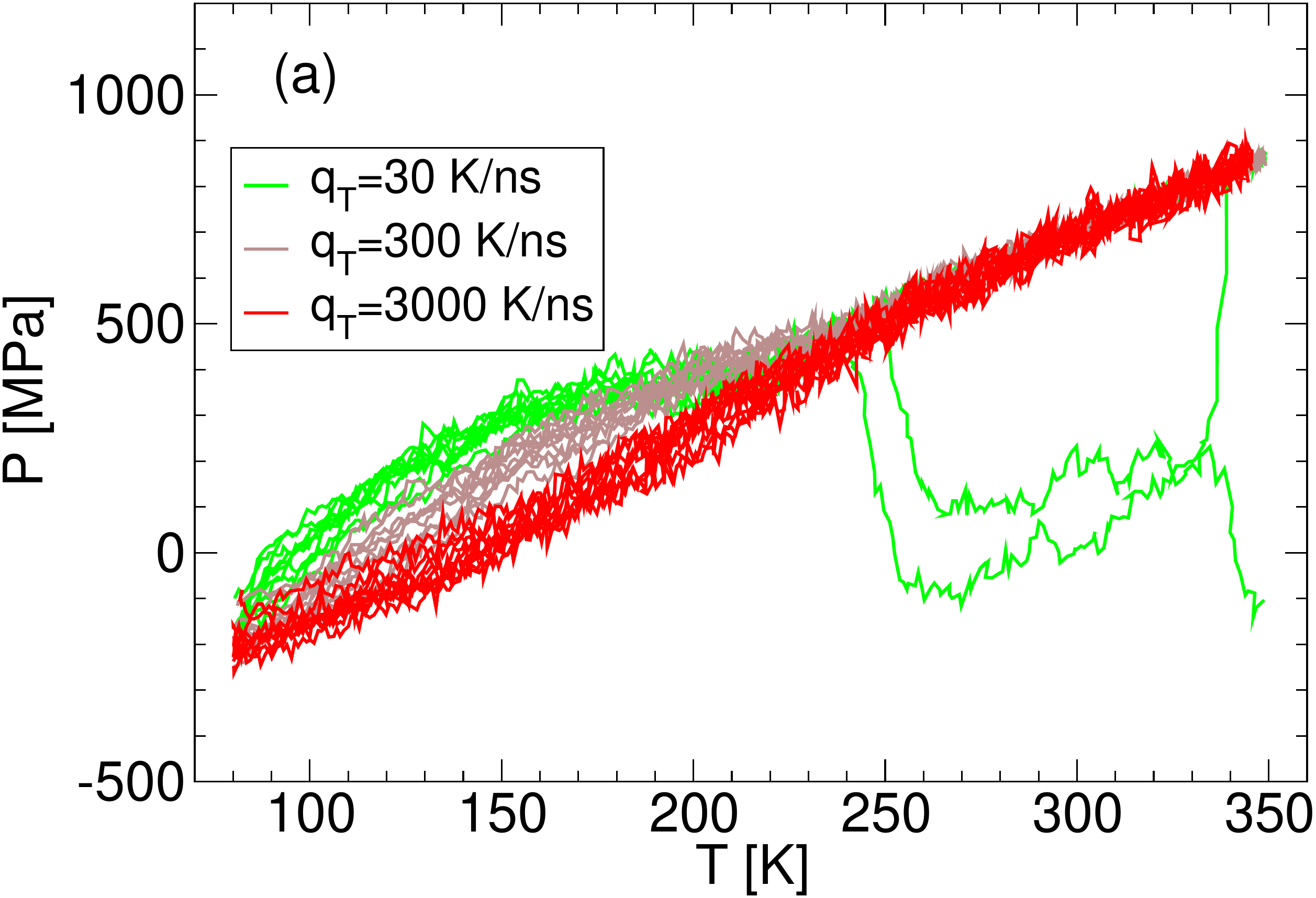}}
\centerline{
\includegraphics[width=7cm]{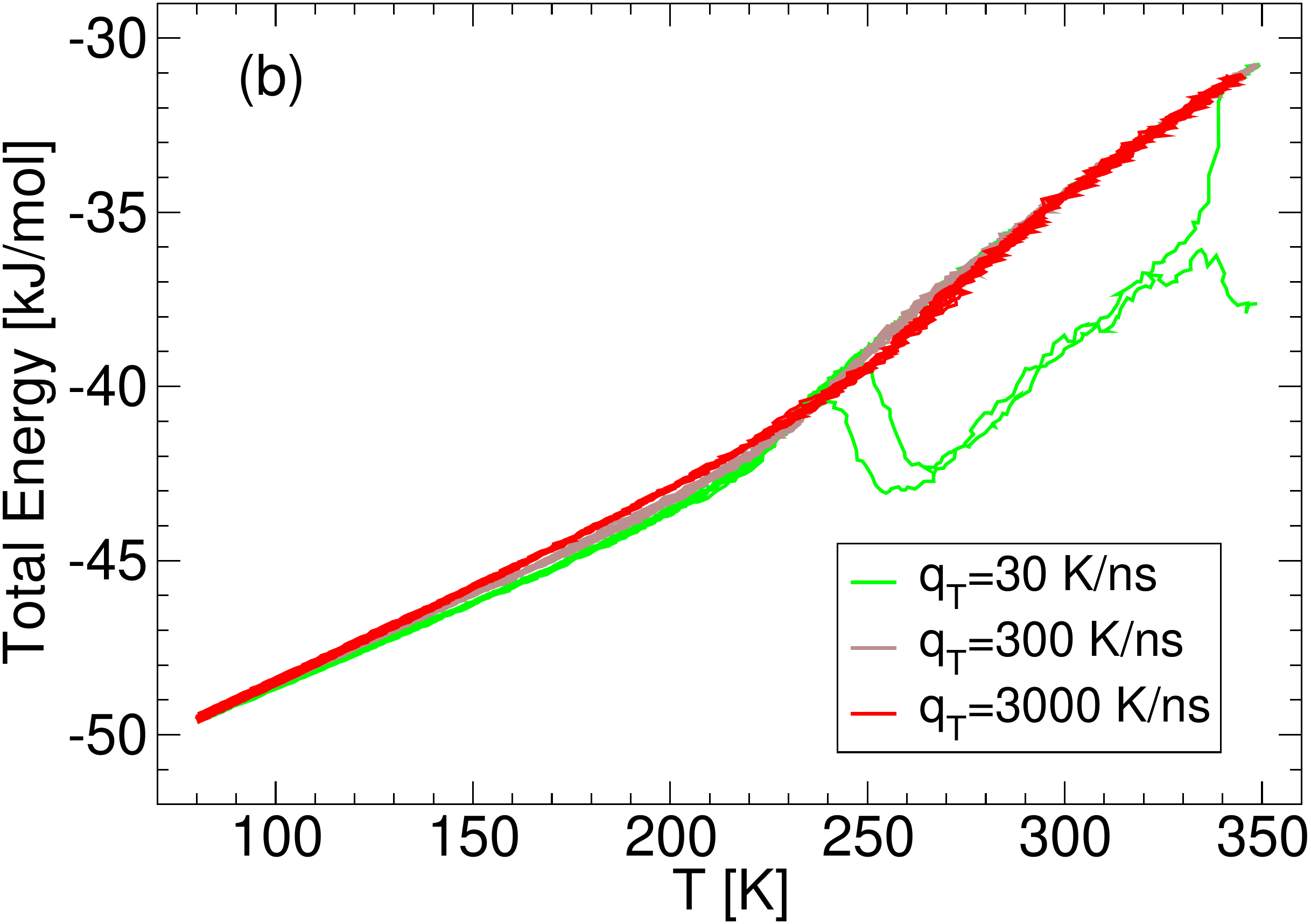}
}
\caption{(a) Pressure and (b) total energy as function of temperature during the isochoric heating of HDA-d at $\rho_i = 1.30$~g/cm$^3$ at different heating rates $q_T$. 
Two runs out of 10 for $q_T = 30$~K/ns (green lines) show crystallization.  All other runs remain in the glass or liquid state.}
\label{fast-heating}
\end{figure}

\begin{figure}      
\centerline{
\includegraphics[width=7cm]{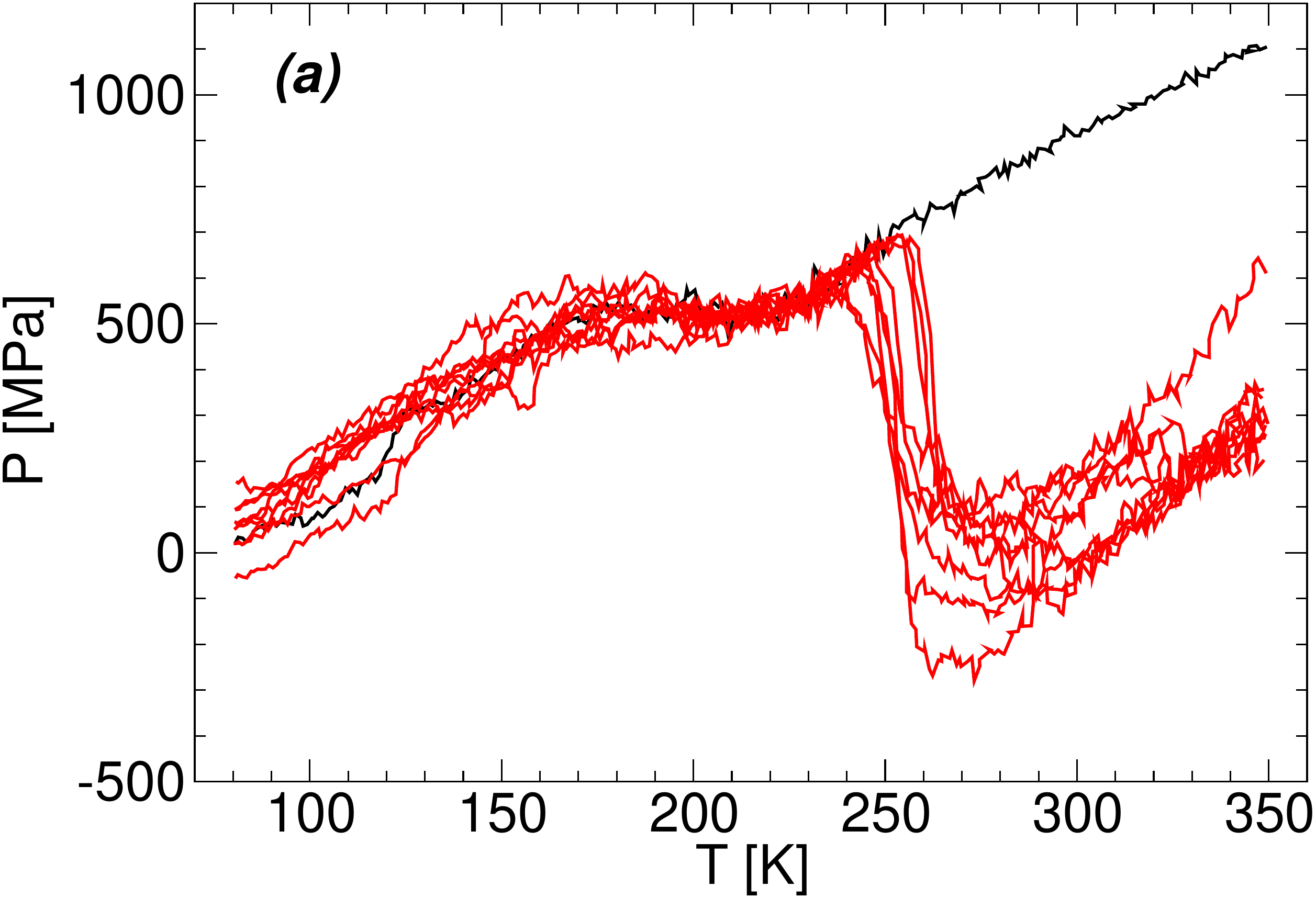}
}
\centerline{
\includegraphics[width=7cm]{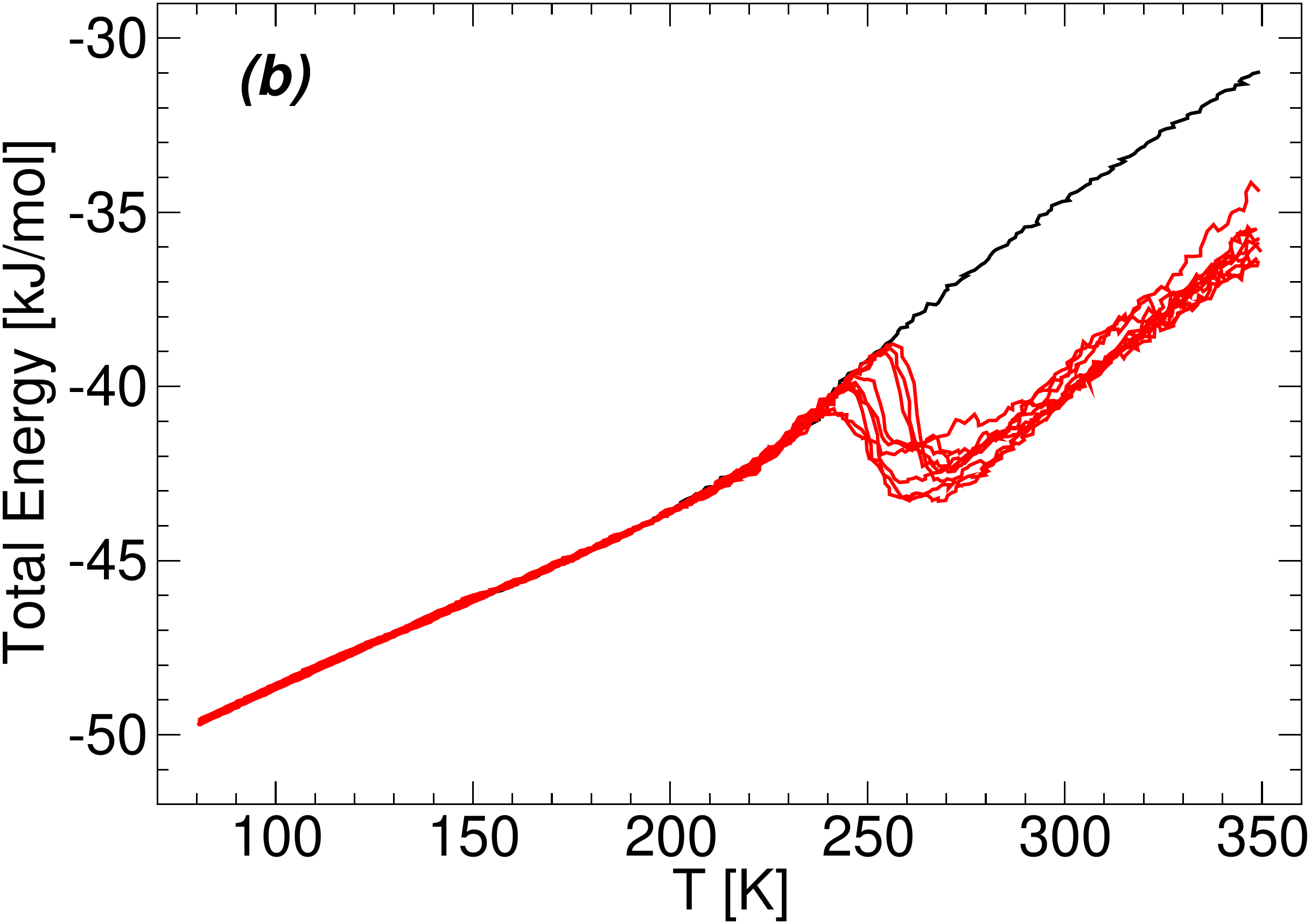}
}
\caption{(a) Pressure and (b) total energy as function of temperature during the isochoric
heating of HDA-d at $\rho_i=1.36$~g/cm$^3$ with $q_T=30$~K/ns (see also magenta
lines in Fig.~\ref{HDAdRates}). The starting HDA-d was produced from LDA with a
compression/decompression rate $q_P=30$~MPa/ns at $T=80$~K.
From the ten independent runs, only one trajectory does not show crystallization
(black line); the rest crystallize at approximately $T>230$~K (red lines).
At low temperatures the system is in the HDA-d state and,
 at the present heating rate, it enters the liquid state at
$T \approx 220$~K. Crystallization occurs randomly at $T_x=240-260$~K (red lines) and
some trajectories briefly reach the liquid state before crystallization occurs.
}
\label{Thermo-HDAd-LIQ_CRSYT}
\end{figure}

\section{Glass Transition and Crystallization Upon Isochoric Heating}
\label{crystalSec}

Next, we discuss briefly the glass transition of HDA-d and crystallization
reported during isochoric heatings.   For simplicity, we focus on the heating runs
discussed in Sec.~\ref{rateSec} above.
The phenomenology is reminiscent to the HDA-to-ice, HDA-to-HDL, and HDA-to-HDL-to-ice
transformations reported in Ref.~\cite{chiu2} upon {\it isobaric} heating.

Fig.~\ref{Thermo-HDAd-LIQ_CRSYT} shows the pressure $P$ and total energy
$E$ as function of temperature upon heating HDA-d at $\rho_i=1.36$~g/cm$^3$.
Crystallization is observed in most of the independent runs (red lines). Only
one trajectory shows no signs of crystallization (black line).
The corresponding mean-square displacement (MSD) of the water O atoms as
function of temperature is shown in Fig.~\ref{msd-HDAd-LIQ_CRSYT}.

At the employed heating rate, $q_T=30$~K/ns, the
MSD is negligible up to $T \approx 220$~K and then starts to increase rapidly
upon further heating.  At $T=250$~K, the MSD reaches approximately $0.6$ to $0.8$~nm$^2$, suggesting
 that molecules have displaced by approximately $0.8$ to $0.9$~nm, i.e., $>2$ times the separation between
OO nearest-neighbors.
 It follows that all samples remain
in the HDA-d state until the glass transition temperature $T_g \approx 220$~K is reached
at $P \approx 550$~MPa.  Again, this temperature
 is very close to the glass transition temperature of HDA obtained
upon {\it isobaric} heating of HDA at $P\approx 550$~MPa and 
$T\approx 225$~K with $q_T=30$~K/ns.
Consistent with the behavior of the MSD, the total energy increases almost linearly with
increasing $T$ for $T<220$~K, while the samples are in the HDA-d state.
As shown in  Fig.~\ref{Thermo-HDAd-LIQ_CRSYT}a, all samples reach the
 same $P$ at $T \approx 200$~K, slightly below $T_g$.

Upon further heating at $T>220$~K, the samples show different behaviors.
When  crystallization is avoided (black line), the MSD and $E$
both increase monotonically with $T$ upon heating, as expected for a system in the
liquid state.  Indeed, as shown in Fig.~\ref{struct-HDAd-LIQ_CRSYT}a,
the OO radial distribution function  (RDF)
of the system at different temperatures shows no indication of crystallization.
Specifically, the RDF reaches the value $\approx 1$ at $r\approx 1$~nm indicating that
there is only short range order in the sample.

In the case where crystallization occurs (red lines), the
  MSD, $E$ and $P$ decrease sharply with $T$ at the crystallization temperature $T_x$.
Not surprisingly, crystallization among the different runs occurs randomly in the range
$T_x=240$ to $260$~K.  Crystallization is confirmed by the RDF of the system;
see Fig.~\ref{struct-HDAd-LIQ_CRSYT}b.
We note that the RDF of the system at $T=260$~K is indeed similar to
that reported in Ref.~\cite{chiu2} for ice VII which was found to form during
isobaric heating of HDA-d at $P>400$~MPa with $q_T=30$~K/ns.
It follows from our discussion that, at the present ultrafast rates, the samples
must reach a HDL-like state over the short temperature interval
($T_g=220$~K)~$<T< (T_x=240$ to $260$~K). Indeed, a similar sequence of transformations from HDA-d to HDL to ice~VII upon isobaric heating was reported in Ref.~\cite{chiu2}. 

Upon further heating, the ultra-fast heating rate employed in this work
forces the crystal to at least partially melt. Specifically, the MSD, $E$ and $P$  increase
sharply with $T$ at $T>300$~K, consistent with a fraction of the system returning to the liquid state.
Fig.~\ref{struct-HDAd-LIQ_CRSYT}b shows that
the structure of the samples retains features of the crystal RDF up to $340$~K, suggesting that melting is not complete in these runs.  Since our heating runs are carried out at constant volume, the system may be in
a state of liquid-ice coexistence.

\begin{figure}      
\centerline{
\includegraphics[width=7cm]{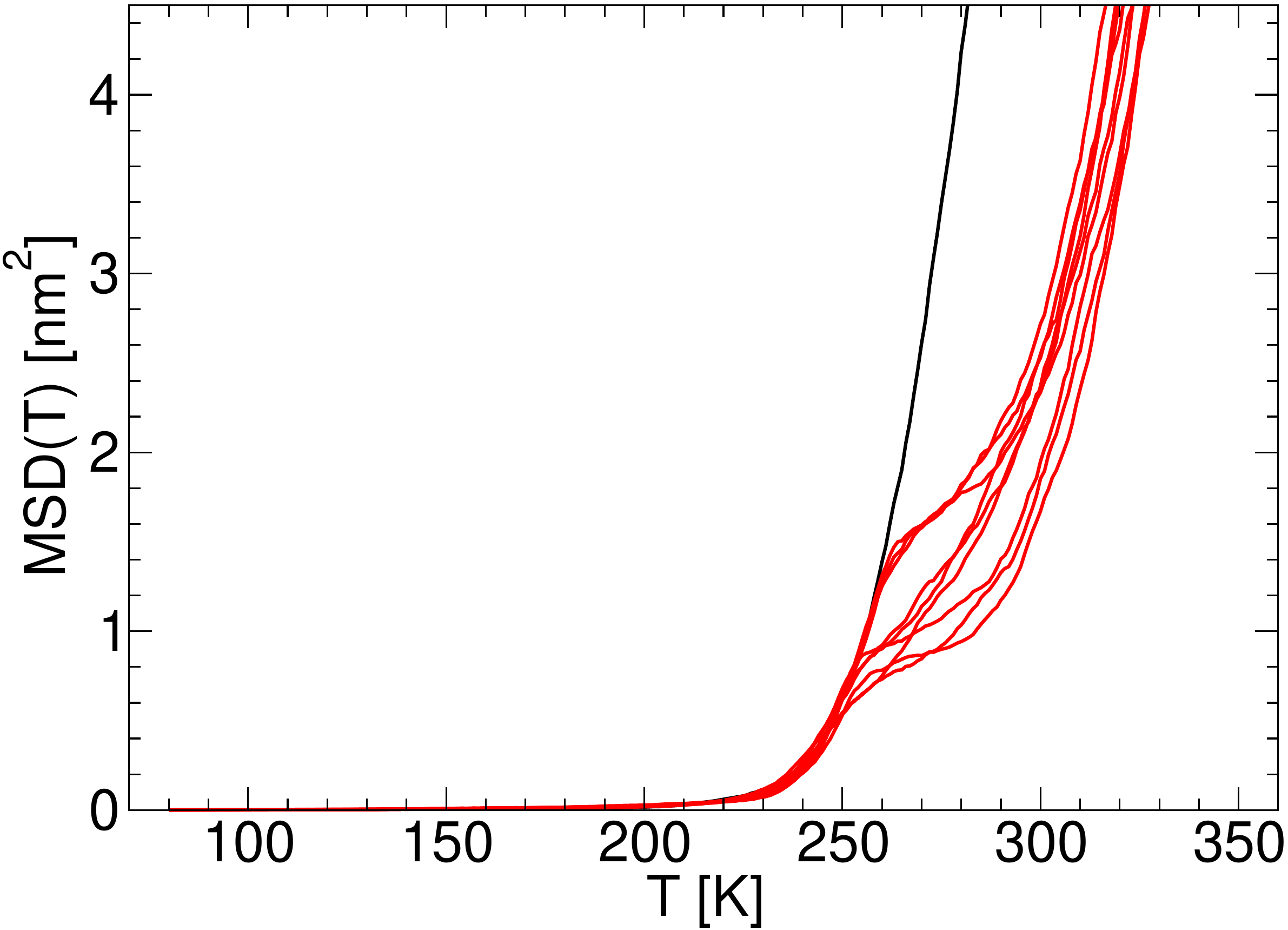}
}
\caption{Mean-square displacement (MSD) of O atoms as function of temperature 
during the isochoric
heating of HDA-d at $\rho_i=1.36$~g/cm$^3$ with $q_T=30$~K/ns; see also Fig.~\ref{Thermo-HDAd-LIQ_CRSYT}.
The MSD is calculated relative to the starting configuration of the system at $T=80$~K.
Molecules do not displace significantly until $T\approx 220$~K and so the system remains in
 the HDA-d state at these temperatures.  At $T\approx 220$~K, the molecules start
to diffuse and the MSD increases rapidly indicating that the system reaches a
 liquid-like state.  Nine out of ten independent runs crystallize at temperatures
 in the range $T=240$ to $260$~K (red lines).
The fast heating rate causes the crystal to subsequently melt.
}
\label{msd-HDAd-LIQ_CRSYT}
\end{figure}

\begin{figure}      
\centerline{
\includegraphics[width=7cm]{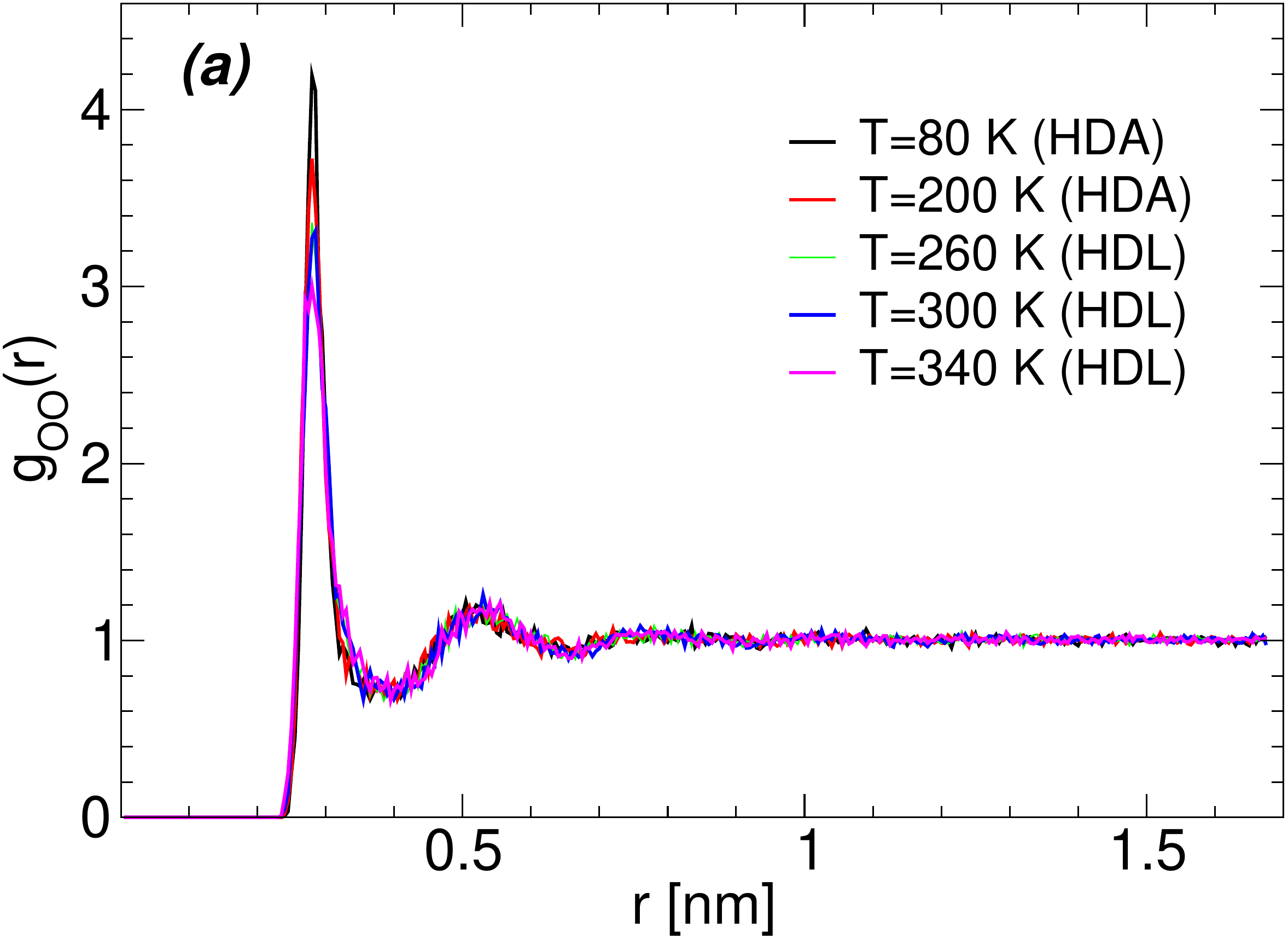}
}
\centerline{
\includegraphics[width=7cm]{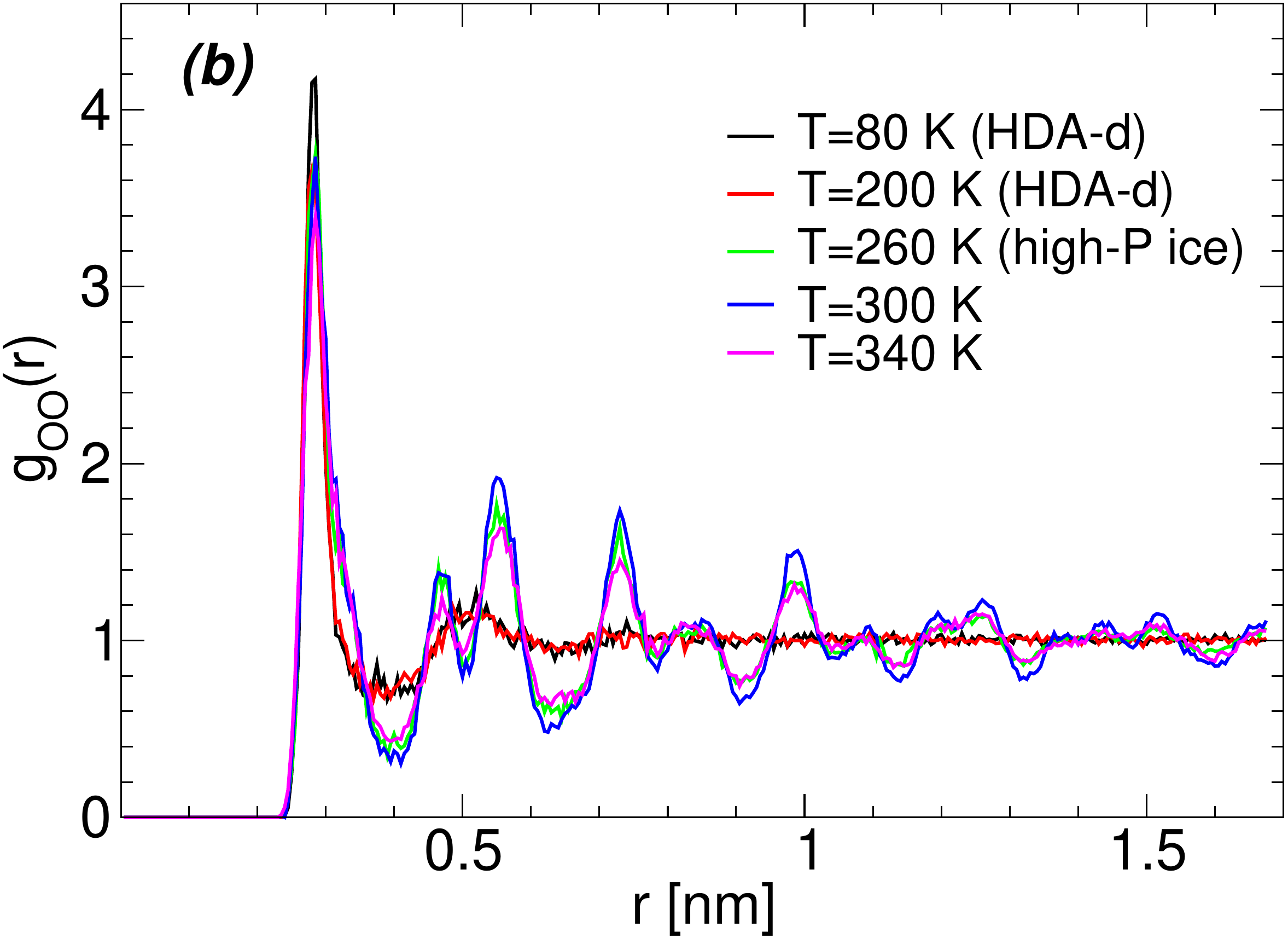}
}
\caption{Oxygen-Oxygen RDFs at selected temperatures upon heating HDA-d at
$\rho_i=1.36$~g/cm$^3$; see also Fig.~\ref{Thermo-HDAd-LIQ_CRSYT}.
(a) RDF of the single sample that does not crystallize upon heating (black line in
Fig.~\ref{Thermo-HDAd-LIQ_CRSYT}). (b) RDF of one of the samples that
 crystallizes upon heating.
}
\label{struct-HDAd-LIQ_CRSYT}
\end{figure}

\begin{figure*}      
\centerline{
\includegraphics[width=8cm]{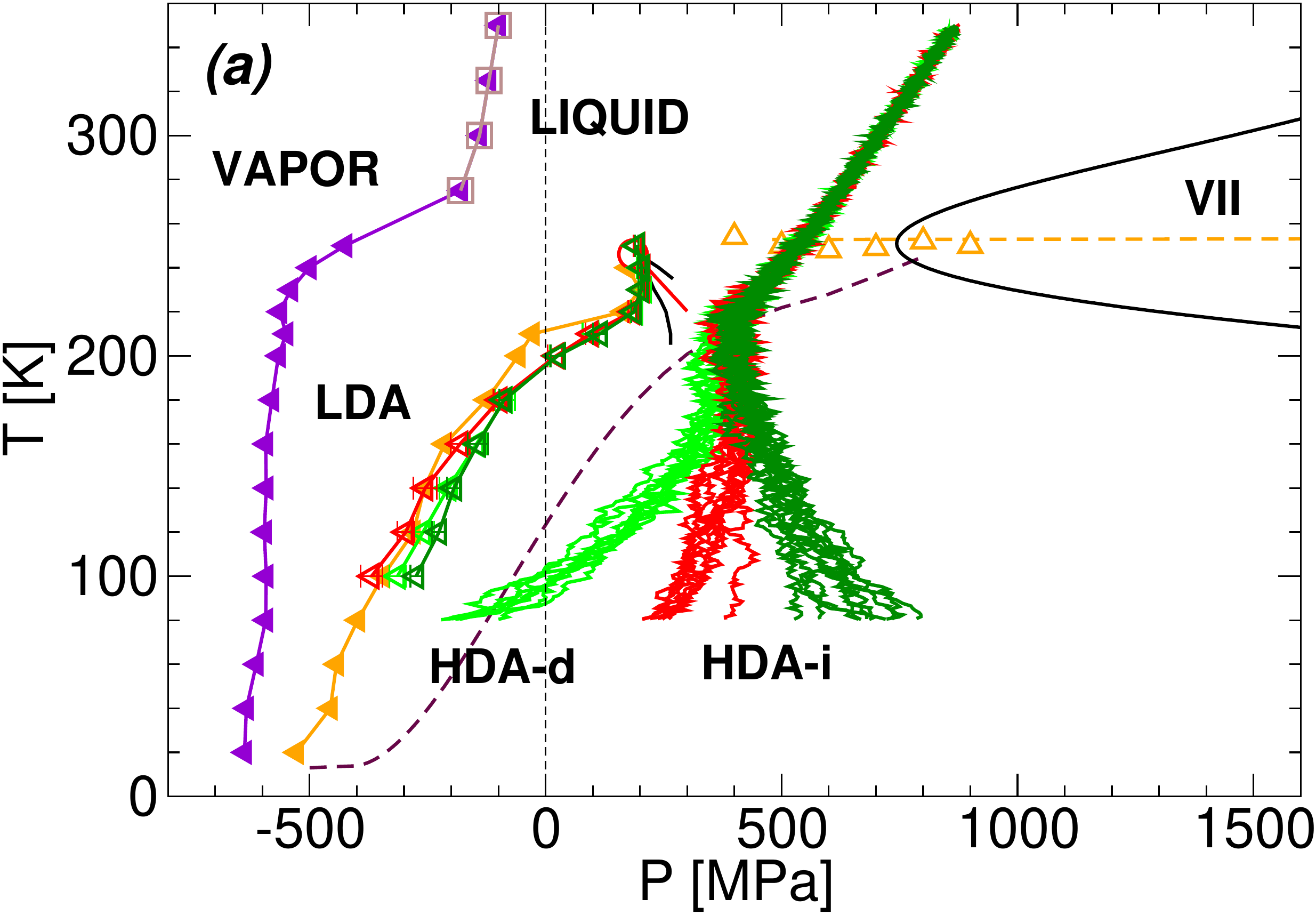}
\includegraphics[width=8cm]{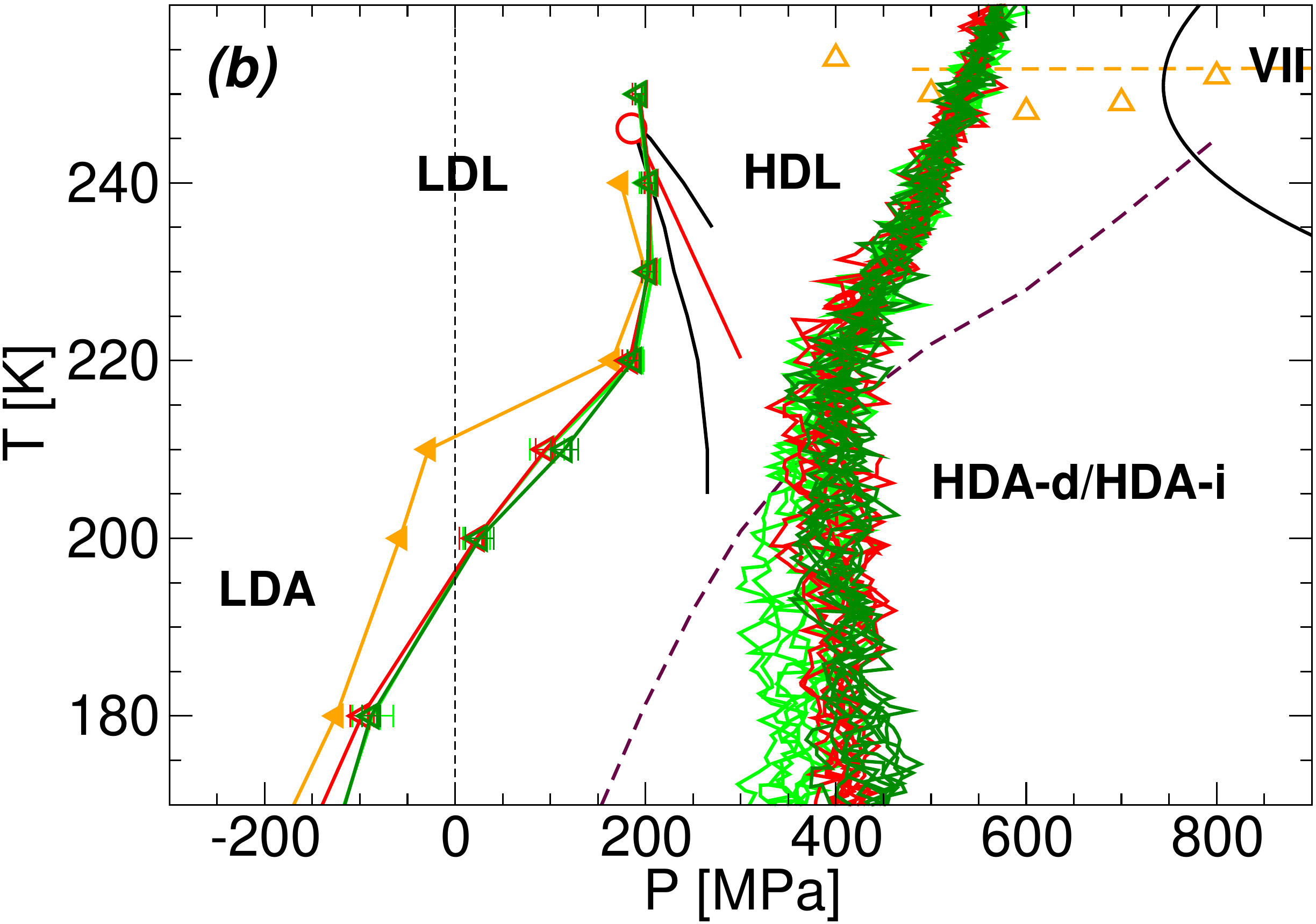}
}
\caption{Same as Fig.~3 of the main manuscript but for the case of HDA-i.  
Temperature as function of pressure upon
heating HDA-i at $\rho_i=1.30$~g/cm$^3$ (red and dark green lines).  Red (dark green)
 lines  correspond to HDA-i obtained by
cooling instantly ($q_T=\infty$) the equilibrium liquid from 
$T_0=220$~K ($T_0=300$~K) to $T=80$~K at constant $P=400$~MPa.  
Empty dark-green and red left-triangles represent the (HDA-i)-to-LDA 
transformation obtained upon isothermal decompression
of HDA-i samples at different tempeartures $T$
(dark-green triangles, for the case $T_0=300$~K, and red triangles, for $T_0=220$~K); 
see also Fig.~\ref{HDAd-HDAi-Decomps}.
For comparison, we also include the density upon heating HDA-d at constant 
$\rho_i=1.30$~g/cm$^3$ [(taken from Fig.~2 of the main manuscript (green lines)] 
and the corresponding (HDA-d)-to-LDA transformation pressure (empty green left-triangles).
At $T \geq 180$~K, the transformation pressures to LDA/LDL
 are independent of whether one starts from HDA-i or HDA-d forms, i.e., the process followed to prepare HDA is not relevant.
Minor differences in the (HDA-i)-to-LDA and (HDA-d)-to-LDA transformation
 pressures are present at $T<180$~K.
}
\label{HDAd-HDAi}
\end{figure*}

\begin{figure}      
\centerline{
\includegraphics[width=7cm]{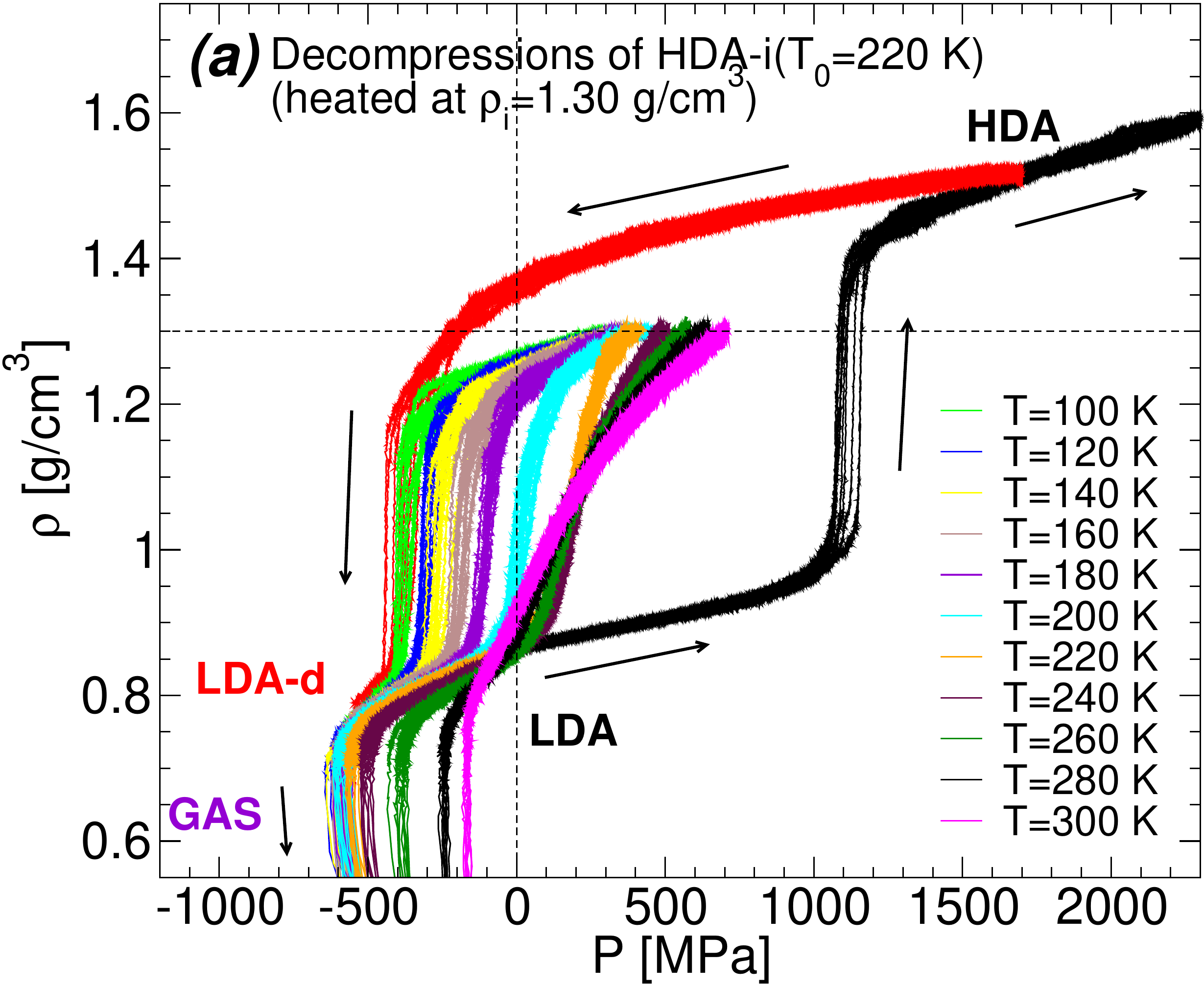}
}
\centerline{
\includegraphics[width=7cm]{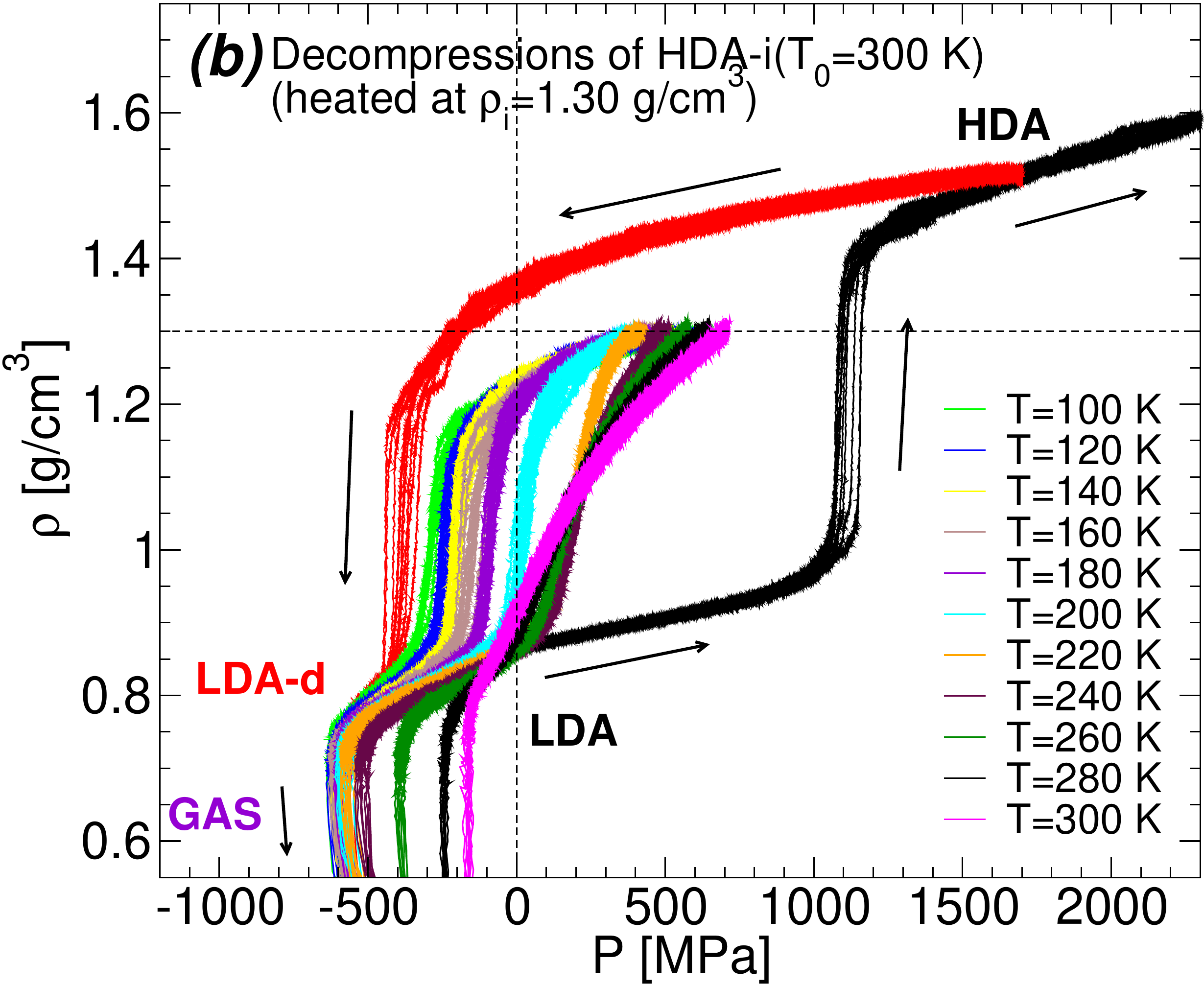}
}
\caption{Density as a function of pressure during the isothermal
decompression of HDA-i
samples heated to different temperatures $T$. The HDA-i samples are prepared
by, first, cooling instantly ($q_T=\infty$) the liquid equilibrated 
at $P=400$~MPa and (a) $T_0=220$~K and (b) $T=300$~K to $T=80$~K.
The HDA-i forms are then decompressed/compressed ($T=80$~K) 
until the density $\rho_i=1.30$~g/cm$^3$ is reached. 
Samples are then heated at constant $\rho_i=1.30$~g/cm$^3$, from $T=80$~K to
the target temperature $T$ with heating rate $q_T=30$~K/ns.
Also included are the pressure-induced LDA-to-HDA and HDA-to-LDA
transformations at $T=80$~K
from Fig.~2b of the main manuscript (black and red lines).
All compression/decompression runs are performed at $q_P=300$~MPa/ns.
No crystallization occurs during decompression of HDA-i ($T_0=300$~K),
while one of the ten decompression runs of 
HDA-i ($T_0=220$~K) exhibits crystallization at $T=240,250,260,260$~K.
Similarly, two runs starting from HDA-d exhibit crystallization
during decompression at $230 \leq T\leq 250$~K.
Decompression trajectories that crystallize are omitted.
}
\label{HDAd-HDAi-Decomps}
\end{figure}

\section{Role of the Process Followed to Prepare Recovered HDA}
\label{HDAiSec}

Glasses, in general, and amorphous ices, in particular, are history-dependent materials.
Accordingly, the process followed in the preparation of the starting HDA sample
could affect the outcome of ultrafast heating-decompression experiments or computer
simulations~\cite{ourST2PEL-2,ourST2PEL-3}.
In this section, we show that our results as presented in the main manuscript 
are robust relative to the method of preparing the
 starting HDA sample at $T=80$~K and $\rho_i$.

In the main manuscript, the starting HDA sample for the heating runs is HDA-d. 
In this section, we use a different HDA form, termed ``HDA-i", which was studied in detail
in Refs.~\cite{chiu1,chiu2,ourST2PEL-1,ourST2PEL-2,ourST2PEL-3}.
Briefly, HDA-i samples are prepared from 
HDL configurations equilibrated at $P=400$~MPa and at temperatures $T_0\ge T_g$.  Here
 we choose $T_0 = 220$ and $300$~K.  After equilibration at $T_0$, the 
HDL samples are 
cooled instantaneously to $T = 80$~K by rescaling the velocities of 
all atoms in the system, producing HDA. That is, these 
HDA-i samples are obtained using an infinite cooling rate.
It follows that the so produced HDA-i samples have the same density and are 
structurally identical to the equilibrium HDL at the starting 
temperature $T_0$ and $P=400$~MPa.
We note that the HDA-i forms obtained from different temperatures $T_0$
can exhibit slightly different structural and thermodynamic 
properties~\cite{ourST2PEL-2,ourST2PEL-3}.
We stress that, since HDA-i is obtained from the equilibrium HDL,
 our HDA-i samples are completely unrelated to our HDA-d samples.

The HDA-i samples produced at $P=400$~MPa and $T=80$~K are 
then isothermally compressed/decompressed at rate $q_P=300$~MPa/ns  at $T=80$~K until they reach 
the density $\rho_i=1.30$~g/cm$^3$.
In the case of $T_0=220$~K [$T_0=300$~K], the pressure of the resulting 
HDA-i samples are within the interval $(200,400)$~MPa [$(525,800)$~MPa].
These HDA-i samples at $\rho_i=1.30$~g/cm$^3$ and $T=80$~K are used as
 starting configurations for the heating runs described in the step (i) of Fig.~1
in the main manuscript.

Fig.~\ref{HDAd-HDAi} shows the trajectory in the $P$-$T$ plane of the HDA-i samples heated at constant 
$\rho_i=1.30$~g/cm$^3$. For comparison, we also include
the trajectories followed by HDA-d samples heated at the same density $\rho_i$, taken from Fig.~1
 (green lines) of the main manuscript.
The pressures of the HDA-i samples at the initial temperature $T=80$~K 
are indeed very different: $P_i=200$ to $400$~MPa for the HDA-i prepared from the liquid 
at $T_0=220$~K (red lines), and $P_i=550$ to $800$~MPa when $T_0=300$~K (dark green lines). 
These pressures are considerably larger than the pressures of HDA-d, which have $-100\leq P \leq -50$~MPa.
It follows that, upon heating, HDA-d, HDA-i ($T_0=220$~K), and HDA-i ($T_0=300$~K) all follow
 different paths within the glass domain. In this glass regime, in the range $T<200$~K, 
the pressure of all the HDA samples varies non-linearly with increasing temperature when $q_T=30$~K/ns.

At $T \approx 200$~K, all HDA forms studied reach the same pressure of $P \approx 400$~MPa.  
This is because all these HDA forms  
have the same density $\rho_i$ and reach the glass transition  at
$T \approx 210$~K  and $P=400$~MPa when $q_T=30$~K/ns.  
Accordingly, HDA-d and HDA-i must transforms to the 
same equilibrium liquid state at $T \approx 210$~K and $P \approx 400$~MPa.
This also explains why 
all heating runs follow a common 
path in the $P$-$T$ plane when $T>210$~K.
In this liquid regime, we find that approximately $P \propto T$.

Despite the different behavior of HDA-d and HDA-i in the glass state, 
the behavior of HDA-d, HDA-i ($T_0=220$~K), and HDA-i ($T_0=300$~K) during isothermal
 decompression is very similar; see Fig.~\ref{HDAd-HDAi-Decomps}.  This similarity is to be expected for samples decompressed at $T>T_g$ for HDA, since these samples all begin the decompression process from the equilibrium HDL phase.  However, the similarity also applies to the decompression runs conducted at $T<T_g$, in which case the HDA-d and HDA-i samples are distinct glasses, with distinct preparation histories.
In particular, we find only very small differences 
in the transformation pressure $P_{\rm H \to L}$.
The $P_{\rm H \to L}$ lines for HDA-d, HDA-i ($T_0=220$~K), and HDA-i ($T_0=300$~K) 
are indicated in Fig.~\ref{HDAd-HDAi} by green, red, and dark-green empty left-triangles, respectively.
Fig.~\ref{HDAd-HDAi} therefore supports the robustness of the results 
reported during the ultrafast heating-decompression pathway considered here, both from computer 
simulations and experiments.

\comment{

}





\end{document}